\documentclass[aps,prd,nofootinbib,12pt]{revtex4}
    \usepackage{graphicx}
    \usepackage{amsmath,amssymb,mathrsfs}
\usepackage{epsf}
\usepackage{slashed}
\usepackage{epsfig}
\usepackage{xcolor}
\allowdisplaybreaks

\newcounter{fig}   \newcommand{\lbfig}[1]{\refstepcounter{fig}
\label{#1} }

\newcommand{\bea}{\begin{eqnarray}}
\newcommand{\eea}{\end{eqnarray}}
\newcommand{\be}{\begin{equation}}
\newcommand{\ee}{\end{equation}}

\newcommand{\re}[1]{(\ref{#1})}

\newcommand{\eqn}{\begin{eqnarray}}
\newcommand{\eqnx}{\end{eqnarray}}

\begin{document}

\title{Kerr black holes with synchronised scalar hair
and boson stars in the Einstein-Friedberg-Lee-Sirlin model}

\author{J.~Kunz}
\affiliation{Carl von Ossietzky University Oldenburg, Germany
Oldenburg D-26111, Germany }
\author{I.~Perapechka}
\affiliation{ Department of Theoretical Physics and Astrophysics,
Belarusian State University, Minsk 220004, Belarus}
\author{Ya.~Shnir}
\affiliation{BLTP, JINR, Dubna 141980, Moscow Region, Russia\\
Department of Theoretical Physics, Tomsk State Pedagogical University, Russia
}

\begin{abstract}
We consider the
Friedberg-Lee-Sirlin model minimally coupled to Einstein gravity
in four spacetime dimensions.
The renormalizable Friedberg-Lee-Sirlin model
consists of two interacting scalar fields, where the mass
of the complex scalar field results from the interaction
with the real scalar field which has a finite vacuum expectation value.
We here study a new family of self-gravitating axially-symmetric,
rotating boson stars in this model.
In the flat space limit these boson stars tend to the
corresponding Q-balls.
Subject to the usual synchronization condition,
the model admits spinning hairy black hole solutions
with two different types of scalar hair.
We here investigate parity-even and parity-odd boson stars
and their associated hairy black holes.
We explore the domain of existence of the solutions
and address some of their physical properties.
The solutions exhibit close similarity to the corresponding boson stars
and Kerr black holes with synchronised scalar hair
in the $O(3)$-sigma model coupled to Einstein gravity
and to the corresponding solutions in the Einstein-Klein-Gordon theory
with a complex scalar field, where the latter are recovered in a limit.
\end{abstract}
\maketitle

\section{Introduction}
The investigation of  self-gravitating scalar field configurations
in 3+1 dimensional asymptotically flat spacetime
has attracted much interest in the last decades.
One of the reasons is that a fundamental scalar field
is a necessary ingredient in models of inflation.
Thus such fields might play an important role in the evolution
of the early Universe.
On the other hand, when scalar fields are present in the Universe,
the gravitational interaction may lead to gravitational collapse
and form localized gravitating objects.
In the case of a complex scalar field
so-called boson stars might arise,
that is, compact, stationary configurations
where the scalar field possesses a harmonic time dependence
\cite{Kaup:1968zz,Ruffini:1969qy}.

Similar static localized field configurations
with finite energy exist in Einstein-Skyrme theory
\cite{Luckock:1986tr,Droz:1991cx,Bizon:1992gb}
and in $\mathrm{SU}(2)$ Einstein-Yang-Mills theory \cite{Bartnik:1988am}
or Einstein-Yang-Mills-Higgs theory
\cite{Lee:1991vy,Breitenlohner:1991aa,Greene:1992fw,Breitenlohner:1994di}.
Certain types of localized gravitating solutions,
like boson stars with appropriate interactions,
or gravitating monopoles, sphalerons and Skyrmions,
are linked to the corresponding flat space solutions,
which represent topological solitons/Q-balls
\cite{Friedberg:1976me,Coleman:1985ki},
or monopoles \cite{tHooft:1974kcl,Polyakov:1974ek},
sphalerons \cite{Klinkhamer:1984di},
and Skyrmions \cite{Skyrme:1961vq},
respectively.

In particular, the Friedberg-Lee-Sirlin model \cite{Friedberg:1976me}
provides an interesting example of a simple
renormalizable two-component scalar field theory with natural interaction terms.
In this model the complex scalar becomes massive due to the coupling
with the real scalar field, since the latter has a finite vacuum expection
value generated via a symmetry breaking potential.
The Q-ball solutions of this model then appear because of
the phase rotation of the complex scalar field,
and the coupling to gravity leads to the respective boson stars.

Gravitating localized solutions of another type
are bound by gravity.
Examples are boson stars without
appropriate self-interactions
\cite{Kaup:1968zz,Ruffini:1969qy},
or the Bartnik-McKinnon solutions \cite{Bartnik:1988am}.
These do not possess a flat space limit.

All these self-gravitating
configurations exist for a certain range of values
of the parameters of the respective theory,
For instance,
there are two branches of self-gravitating Skyrmions
\cite{Droz:1991cx,Bizon:1992gb},
where the lower in energy branch is linked to the flat space Skyrmion
in the limit of a vanishing effective gravitational coupling.
This lower branch of solutions then ends at some critical maximal value
of the gravitational coupling,
where it bifurcates with the second, higher in energy branch,
which extends all the way backwards to the limit of zero coupling.
Likewise, for gravitating monopoles the
lower in energy branch is linked to the flat space monopole,
however, the second branch ends
when an extremal Reissner-Nordstr\"om configuration is reached
(in the exterior)
\cite{Lee:1991vy,Breitenlohner:1991aa}.

Boson star configurations, on the other hand,
possess a presumably infinite number of
branches, representing an inspiraling of the solutions
towards a limiting solution,
when the dependence of the mass or particle number
on the frequency or on the radius is considered
\cite{Friedberg:1986tp,Friedberg:1986tq,Kleihaus:2005me,Kleihaus:2007vk}.
This spiraling behavior is reminiscent of the mass radius
relation of neutron stars beyond the maximum mass star.
For other physical quantities this translates
into an oscillating behavior for boson stars and neutron stars, alike.

Notably, many of those regular particle-like gravitating solitons,
like the gravitating monopoles, sphalerons and Skyrmions
or the Bartnik-McKinnon solutions, can be linked
to hairy black holes in the limit of vanishing event horizon radius
\cite{Luckock:1986tr,Volkov:1989fi}.
These static black holes provide the first known counter-examples to the
celebrated no-hair conjecture \cite{Ruffini:1971bza}
(see also \cite{Volkov:1998cc,Herdeiro:2015waa,Volkov:2016ehx}
for further references and discussion).

In contrast, the spherically symmetric boson star solutions
\cite{Kaup:1968zz,Ruffini:1969qy,Friedberg:1986tp,Friedberg:1986tq}
cannot be generalized to contain a small Schwarzschild black hole
in their inner region.
More general, it has been shown that there is no regular static
asymptotically flat solution with an event horizon in these models
with a complex scalar field, which harmonically depends on time
\cite{Pena:1997cy,Hod:2018dij}.
This situation is not unique, however.
For example, regular static
spherically symmetric self-gravitating solutions
of the generalized Skyrme model \cite{Adam:2016vzf,Gudnason:2016kuu}
terminate at a singular solution,
and cannot be continuously connected to a
static hairy black hole \cite{Perapechka:2016cof}.

An interesting aspect for all such gravitating regular
and black hole solutions is their possible generalization
to include rotation, since rotation is ubiquitous in
the Universe.
For instance, for boson stars there is no slow rotating limit
\cite{Kobayashi:1994qi},
however, they can rotate rapidly
\cite{Schunck:1996he,Ryan:1996nk,Yoshida:1997qf}.
Other rotating regular configurations include
the gravitating Skyrmions and
\textit{Q-clouds} \cite{Ioannidou:2006nn},
gravitating dyons and vortex rings \cite{Kleihaus:2005fs,Kleihaus:2007vf}
or the spinning topological solitons of the non-linear
$\mathrm{O}(3)$ sigma model \cite{Herdeiro:2018djx}.

Whereas various rotating hairy black holes were obtained before
\cite{Kleihaus:2000kg,Kleihaus:2002ee,Kleihaus:2003sh,Kleihaus:2003df,Kleihaus:2004gm,Kleihaus:2016rgf},
only recently a spinning complex scalar field
was considered in a Kerr black hole spacetime,
first perturbatively and then with back reaction
\cite{Hod:2012px,Herdeiro:2014goa,Herdeiro:2014jaa,Herdeiro:2015gia,Hod:2014baa,Benone:2014ssa,Herdeiro:2014pka,Herdeiro:2014ima}.
In that case, the rotating hairy black holes obey a synchronization
condition between the angular velocity of the event horizon
and a phase frequency of the scalar field.
Numerous hairy black holes obeying such a synchronization condition
have been studied by now. Further examples are given in
\cite{Kleihaus:2015iea,Herdeiro:2015tia,Herdeiro:2015kha,Herdeiro:2016tmi,Brihaye:2016vkv,Hod:2017kpt,Herdeiro:2017oyt,Herdeiro:2018daq,Herdeiro:2018djx,Wang:2018xhw,Delgado:2019prc,Kunz:2019bhm}.

Interestingly, the Q-balls in the Friedberg-Lee-Sirlin model
in flat space may also exist in the limiting
case of vanishing scalar potential \cite{Levin:2010gp,Loiko:2018mhb}.
In this limit the real component of the
scalar field becomes massless, thus it possesses
a Coulomb-like asymptotic tail.
This leads to the interesting question,
as to whether one can find similar rotating self-gravitating
asymptotically flat solutions,
which are either regular or possess an event horizon.
Such black hole solutions would represent a new type of hairy black holes,
quite different from the scalarized hairy black holes
of scalar-tensor theory,
where the real scalar field is associated with
the gravitational interaction \cite{Kleihaus:2015iea}.

In this paper we answer this question positively.
In particular, we extend the study of the spinning boson stars and
hairy black holes, by constructing new families of stationary rotating
solutions in the Friedberg-Lee-Sirlin model
minimally coupled to Einstein gravity.
The boson star solutions possess properties
that are similar to those of the rotating boson stars
with a single complex scalar field
\cite{Schunck:1996he,Ryan:1996nk,Yoshida:1997qf},
featuring both parity-even and parity-odd configurations
\cite{Kleihaus:2005me,Kleihaus:2007vk}.
In the flat space limit, they are linked to
the corresponding spinning flat-space Q-balls
\cite{Volkov:2002aj,Kleihaus:2005me,Kleihaus:2007vk,Radu:2008pp,Brihaye:2008cg,Loiko:2018mhb}.

All these rotating boson star solutions are related to rotating hairy
black hole solutions,
where the phase rotation of the massive complex component
is synchronized with the angular velocity of the horizon.
An important novelty of these solutions is the fact that
in the limit of vanishing potential these solutions
correspond to black holes with {\it two} different types of scalar hair,
where the second real scalar field is massless.
Further, in the opposite limit of infinite mass of the real component,
the solutions of the model effectively reduce to the corresponding
boson stars and hairy black holes of the Einstein-Klein-Gordon model
\cite{Herdeiro:2014goa,Herdeiro:2015gia,Herdeiro:2015waa,Kunz:2019bhm}.
We here show that, depending on the values of the
parameters of the model and, in particular, the horizon radius parameter,
the branch structure of the solutions varies from the typical
spiral pattern of boson stars
\cite{Friedberg:1986tp,Friedberg:1986tq,Kleihaus:2005me,Kleihaus:2007vk,Tamaki:2010zz,Collodel:2017biu},
to a simpler branch structure known for various types of hairy black holes
\cite{Droz:1991cx,Bizon:1992gb,
Kleihaus:2015iea,Adam:2016vzf,Gudnason:2016kuu,
Kleihaus:2000kg,Kleihaus:2002ee,Kleihaus:2003sh,Kleihaus:2003df,Kleihaus:2004gm,
Herdeiro:2014goa,Herdeiro:2014jaa,Herdeiro:2015gia,Kleihaus:2015iea,Herdeiro:2015tia,Herdeiro:2015kha,Herdeiro:2016tmi,Brihaye:2016vkv,Wang:2018xhw,Delgado:2019prc,Kunz:2019bhm,Perapechka:2017bsb,Herdeiro:2018daq,Herdeiro:2018djx}.

\section{The Model}
We consider the 3+1 dimensional action
\be
\label{action}
S=\int d^4x \sqrt{-g} \left(\frac{R}{4 \alpha^2} - \mathcal{L}_{m} \right)\, ,
\ee
where the gravity part  is the usual Einstein-Hilbert action,
$\alpha^2=4\pi G$ is the gravitational coupling,
$R$ is the curvature scalar and $G$ is Newton's constant.
The Lagrangian of the matter fields $\mathcal{L}_{m}$
is given by the two-component Friedberg-Lee-Sirlin model
\cite{Friedberg:1976me}
\be
\label{lagFLS}
\mathcal{L}_\mathrm{m} =
\frac{1}{2}\left(\partial_\mu\psi\right)^2+\left|\partial_\mu\phi\right|^2 +
m^2\psi^2|\phi|^2-\mu^2\left(\psi^2 - v^2\right)^2\, .
\ee
Here a real self-interacting scalar field $\psi$ is coupled to a
complex scalar field $\phi$, the parameters
$m$ and $\mu$ are the real positive coupling constants and $v$ is
the vacuum expectation value of the real scalar field $\psi$.
The first two parts in \re{lagFLS} are the
usual kinetic terms for the real and complex field, respectively,
the third is the interaction term, and the last
term gives the potential of the real scalar field.

The potential is chosen such that in the vacuum $\psi \to v$,
and the complex field $\phi$ becomes massive with mass $m v$
due to the coupling with its real partner.
When expanded around its vacuum expectation value $v$,
the fluctuations of the real scalar field are associated with
a mass $\sqrt{8} \mu v$, thus $\mu$ represents a mass parameter
for the real scalar field.
In the limit of vanishing mass parameter $\mu \to 0$ but fixed
vacuum expectation value $v$, the real scalar field
becomes massless and thus long-ranged.
The complex component $\phi$ still acquires mass in this limit
due to the coupling with the Coulomb-like field $\psi$.

Note that two of the four parameters of the model \re{action}
can be rescaled away via transformations of the coordinates and the fields,
\be
\alpha \to v \alpha\, , \quad x_\mu \to m v x_\mu\, ,
\quad \psi \to \frac{\psi}{v}\, , \quad \phi \to \frac{\phi}{v} \, .
\ee
After such a rescaling $\tilde \alpha = v \alpha$ and $\tilde \mu = \mu/m$
will be the remaining parameters. However, we will only set $v=1$
in the following, retaining $\tilde \alpha$ (omitting the tilde),
$m$ and $\mu$.

The model \re{action} is invariant under global $\mathrm{U}(1)$
transformations of the complex field
$\phi\to\phi e^{\delta}$, where the parameter $\delta$ is a constant.
The following Noether current is associated with this symmetry,
\be
\label{Noether}
j_\mu = i(\phi\partial_\mu\phi^\ast-\phi^\ast\partial_\mu\phi) \, ,
\ee
with the corresponding charge $Q=\int{\sqrt{-g}j^t d^3 x}$.

Variation of the action \re{action} with respect to the metric
leads to the Einstein equations
\be
\label{Einstein}
R_{\mu\nu}-\frac{1}{2}Rg_{\mu\nu}=2\alpha^2 T_{\mu\nu},
\ee
where
\be
\label{SET}
T_{\mu\nu}=\partial_\mu\psi\partial_\nu\psi+\left(\partial_{\mu}\phi\partial_{\nu}\phi^\ast
+\partial_{\nu}\phi\partial_{\mu}\phi^\ast\right) -\mathcal{L}_\mathrm{m}g_{\mu\nu}
\ee
is the stress-energy tensor of the scalar fields.

The corresponding equations of motion of the scalar fields read
\be
\label{scaleq}
\begin{split}
    \Box \psi&=2\psi\left(m^2|\phi|^2+2\mu^2\left(1-\psi^2\right)\right),\\
    \Box \phi&=m^2\psi^2\phi \, ,
\end{split}
\ee
where $\Box$ represents the covariant d'Alembert operator.
It follows from the linearized field equations \re{scaleq}
that the parameters $\mu$ and $m$ indeed determine the mass
of the real and complex scalar fields, respectively.
Notably, the flat-space localized regular solutions
of the Friedberg-Lee-Sirlin model \re{lagFLS}
exist in the limit of vanishing scalar potential, $\mu \to 0$,
when the vacuum expectation value of the real component $\psi$ is kept non-zero
\cite{Levin:2010gp,Loiko:2018mhb}.
They represent Q-balls with a long-range massless scalar component.
These Q-balls are similar to those
of the Wick-Cutkosky model \cite{Wick-Cutkosky}
revisited recently in
\cite{Nugaev:2016uqd,Panin:2018uoy}.

In the opposite limit, $\mu \to \infty$,
the real component of the model \re{lagFLS} trivializes, $\psi =1$,
and the massive complex field $\phi$ satisfies the Klein-Gordon equation.
Clearly, spatially localized stationary spinning solutions
of this equation do not exist in the flat space.
However, there are families of corresponding boson stars
and hairy black holes with synchronised hair
in the complex-Klein-Gordon field theory
minimally coupled to Einstein's gravity \cite{Herdeiro:2014goa,Herdeiro:2015gia,Herdeiro:2015waa,Wang:2018xhw,Kunz:2019bhm}.

\section{Spinning axially-symmetric configurations}
\subsection{Stationary axially symmetric ansatz and boundary conditions}
Spherically symmetric self-gravitating regular solutions
of the equations \re{Einstein},\re{scaleq}
do not admit
spherically symmetric generalizations with a horizon \cite{Pena:1997cy}.
In the present paper we shall consider spinning regular and
hairy black hole solutions to the system \re{Einstein},\re{scaleq}.
We note, that spinning self-gravitating regular solutions of
an extended model with two complex interacting scalar fields
were considered before \cite{Kunz:2013wka}.

To obtain stationary spinning axially-symmetric solutions
we take into account the presence of two commuting
Killing vector fields $\xi=\partial_t$ and $\eta=\partial_\varphi$,
where  $t$ and $\varphi$ are the time and azimuthal coordinates, respectively.
In these coordinates the metric can be written
in isotropic coordinates in the Lewis-Papapetrou form
\be
\label{metrans}
ds^2=-F_0 dt^2 +F_1\left(dr^2+r^2 d\theta^2\right)+ r^2\sin^2 \theta F_2  \left(d\varphi-\frac{W}{r} dt\right)^2\, ,
\ee
where the four metric functions $F_0, F_1, F_2$ and $W$
depend on $r$ and $\theta$ only.

For the scalar fields we adopt the axially-symmetric ansatz
\be
\label{scalans}
\psi=X(r,\theta),\quad \phi=Y(r,\theta)e^{i\omega t+n\varphi}\, ,
\ee
where the real profile functions $X$ and $Y$ depend on
the radial coordinate $r$ and the polar angle $\theta$,
the frequency of the spinning complex field $\omega$
is a parameter of the model,
and $n\in\mathbb{Z}$ is the azimuthal winding number,
also referred to as rotational quantum number.
For stationary spherically symmetric configurations $n = 0$,
and the system of equations \re{Einstein},\re{scaleq} reduces
to the set of equations of the corresponding boson stars.

To obtain hairy black holes,
we assume the existence of a rotating event horizon,
located at a constant value of the radial variable, $r=r_h>0$.
The Killing vector of the horizon is the helicoidal vector field
\be
\label{Killingrh}
\chi=\xi +\Omega_h \eta\, ,
\ee
where the horizon angular velocity $\Omega$
is fixed by the value of the metric function $W$ on the horizon
$$
\Omega_h =  -\frac{g_{\phi t}}{g_{tt}}\biggl.\biggr|_{r=r_h}
= W\biggl.\biggr|_{r=r_h} \, .
$$
The presence of a rotating horizon allows
to form stationary scalar clouds,
supported by the synchronisation condition
\cite{Hod:2012px,Herdeiro:2014goa,Herdeiro:2015gia,Herdeiro:2015waa,Herdeiro:2018djx}
\be
\label{synchron}
\omega = n \Omega_h \,
\ee
between the event horizon angular velocity $\Omega_h$,
the complex scalar field frequency $\omega$, and the winding number $n$.
This condition implies
that there is no flux of the complex scalar field into the black hole.

It is convenient to make use of the exponential parametrization
of the metric fields
\be
\label{subst}
F_0=\frac{\left(1-\frac{r_h}{r}\right)^2}{\left(1+\frac{r_h}{r}\right)^2}e^{f_0},
\quad F_1=\left(1+\frac{r_h}{r}\right)^4 e^{f_1},
\quad F_2=\left(1+\frac{r_h}{r}\right)^4 e^{f_2}\, ,
\ee
where the functions $f_i$ depend on the radial coordinate $r$
and the polar angle $\theta$.
Then a power series expansion near the horizon
yields the following conditions of regularity for the profile functions
$X, Y$ and the metric functions $f_i$:
\be
\label{bchor}
\partial_r X\bigl.\bigr|_{r=r_h}=\partial_r Y\bigl.\bigr|_{r=r_h}=\partial_r f_0\bigl.\bigr|_{r=r_h}=\partial_r f_1\bigl.\bigr|_{r=r_h}
=\partial_r f_2\bigl.\bigr|_{r=r_h} = 0\, .
\ee
These Neumann boundary conditions must be supplemented
by the synchronization condition \re{synchron}
imposed on the metric function $W$.

Requirement of asymptotic flatness implies that, as $r\to \infty$,
the metric approaches the Minkowski limit,
and the scalar fields are taking their vacuum values
\be
\label{bcinf}
X\bigl.\bigr|_{r\to \infty}=1\, , \quad
Y\bigl.\bigr|_{r\to \infty}=f_0\bigl.\bigr|_{r\to \infty}
=f_1\bigl.\bigr|_{r\to \infty}=f_2\bigl.\bigr|_{r\to \infty}
=W\bigl.\bigr|_{r\to \infty}=0\, .
\ee
Demanding axial symmetry and regularity imposes
the following boundary conditions
on the symmetry axis for $\theta=0,\pi$
\be
\label{bcpole}
\partial_\theta X\bigl.\bigr|_{\theta = 0,\pi} =
Y\bigl.\bigr|_{\theta = 0,\pi} =
\partial_\theta f_0\bigl.\bigr|_{\theta = 0,\pi} =
\partial_\theta f_1\bigl.\bigr|_{\theta = 0,\pi} =
\partial_\theta f_2\bigl.\bigr|_{\theta = 0,\pi} =
\partial_\theta W\bigl.\bigr|_{\theta = 0,\pi}=0\, .
\ee
We also require the solutions to be $\mathbb{Z}_2$-symmetric
with respect to reflection symmetry $\theta \to \pi -\theta$
in the equatorial plane $\theta = \pi/2$.
Thus, we can restrict the range of values
of the angular variable as $\theta \in [0,\pi/2]$.
The corresponding boundary conditions on the equatorial plane are
\be
\label{bcaxis}
\partial_\theta X\bigl.\bigr|_{\theta = \frac{\pi}{2}} =
\partial_\theta f_0\bigl.\bigr|_{\theta = \frac{\pi}{2}} =
\partial_\theta f_1\bigl.\bigr|_{\theta=\frac{\pi}{2}}
= \partial_\theta f_2\bigl.\bigr|_{\theta=\frac{\pi}{2}}
= \partial_\theta W\bigl.\bigr|_{\theta=\frac{\pi}{2}}=0 \, .
\ee
\begin{figure}[hbt]
    \begin{center}
        \includegraphics[width=.48\textwidth, trim = 40 20 90 20, clip = true]{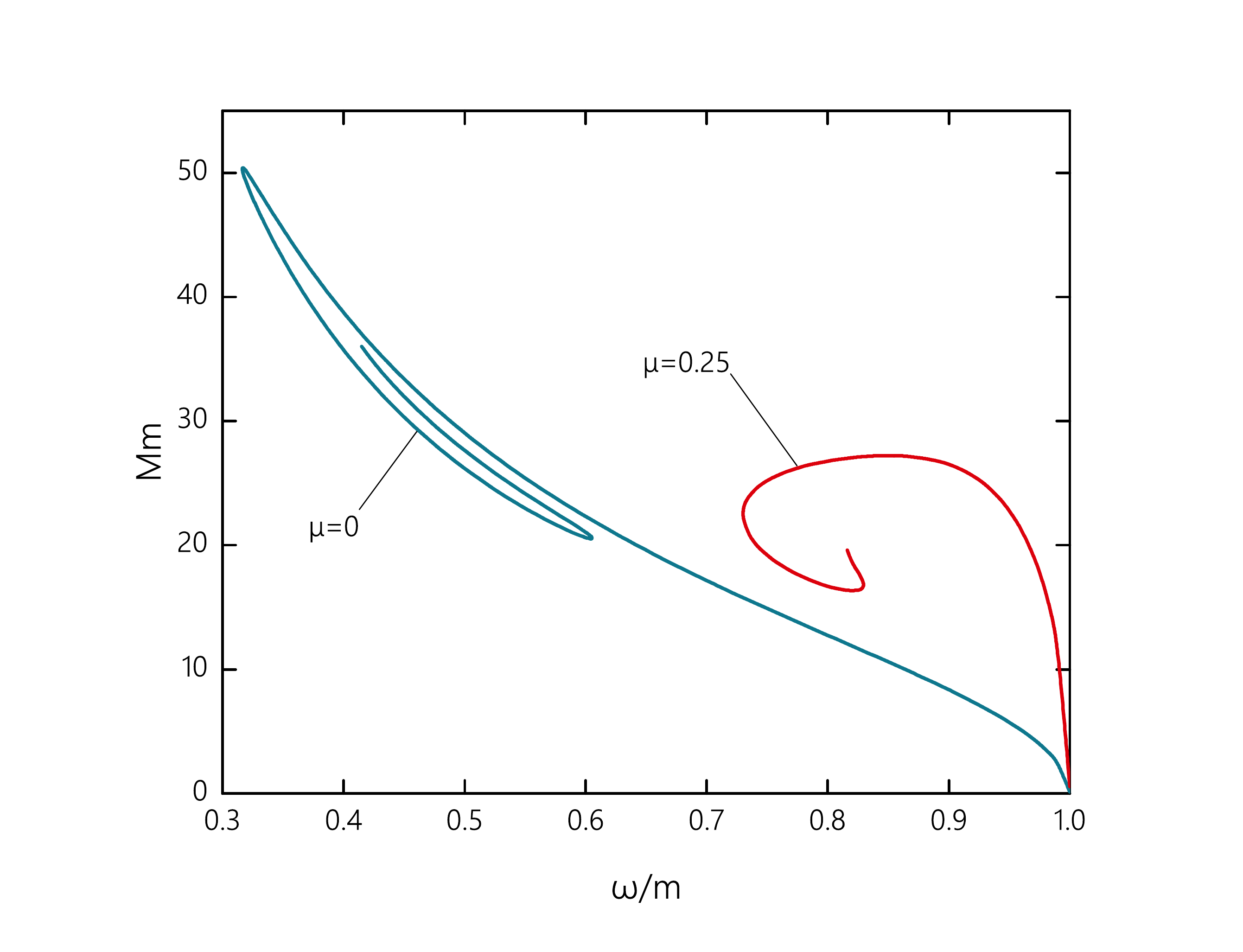}
        \includegraphics[width=.48\textwidth, trim = 40 20 90 20, clip = true]{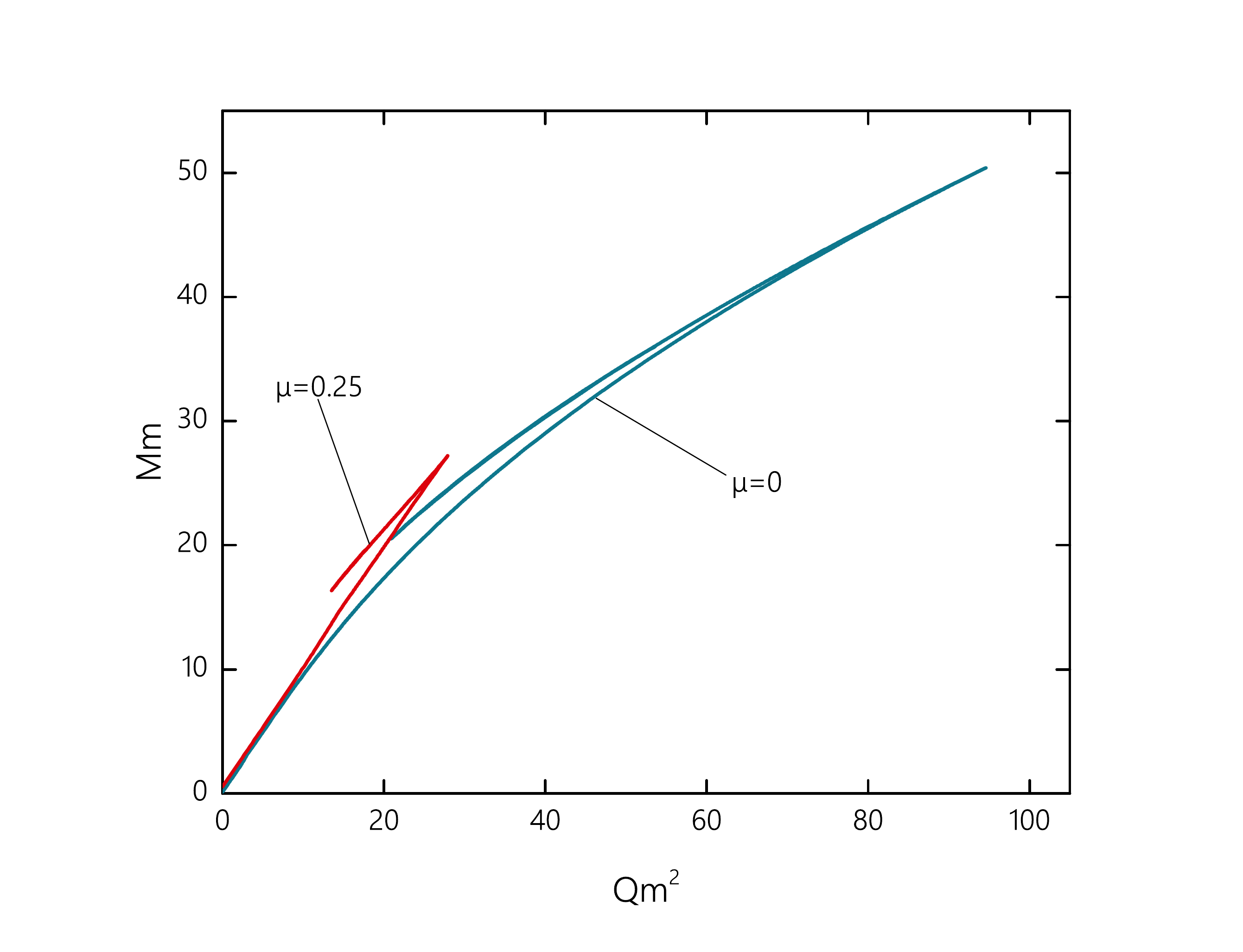}
    \end{center}
    \caption{\small
        The ADM mass $M$
        vs the angular frequency $\omega$ (left plot) and Noether charge $Q$ (right plot)
at $\mu/m = 0.25$,  $ \mu=0$ and $\alpha = 0.5$ for parity-even
boson stars.}
    \lbfig{M_n0}
\end{figure}

Further, there are two different types of axially symmetric solutions,
which possess different parity of the complex scalar field
\cite{Kleihaus:2005me,Kleihaus:2007vk,Loiko:2018mhb}.
The corresponding boundary conditions on the complex component of the field in the equatorial plane are
$\partial_\theta Y\bigl.\bigr|_{\theta = \frac{\pi}{2}} = 0$
for parity-even solutions,
and $Y\bigl.\bigr|_{\theta = \frac{\pi}{2}} = 0$ for parity-odd configurations.

\begin{figure}[p!]
    \begin{center}
        \includegraphics[width=.48\textwidth, trim = 40 20 90 20, clip = true]{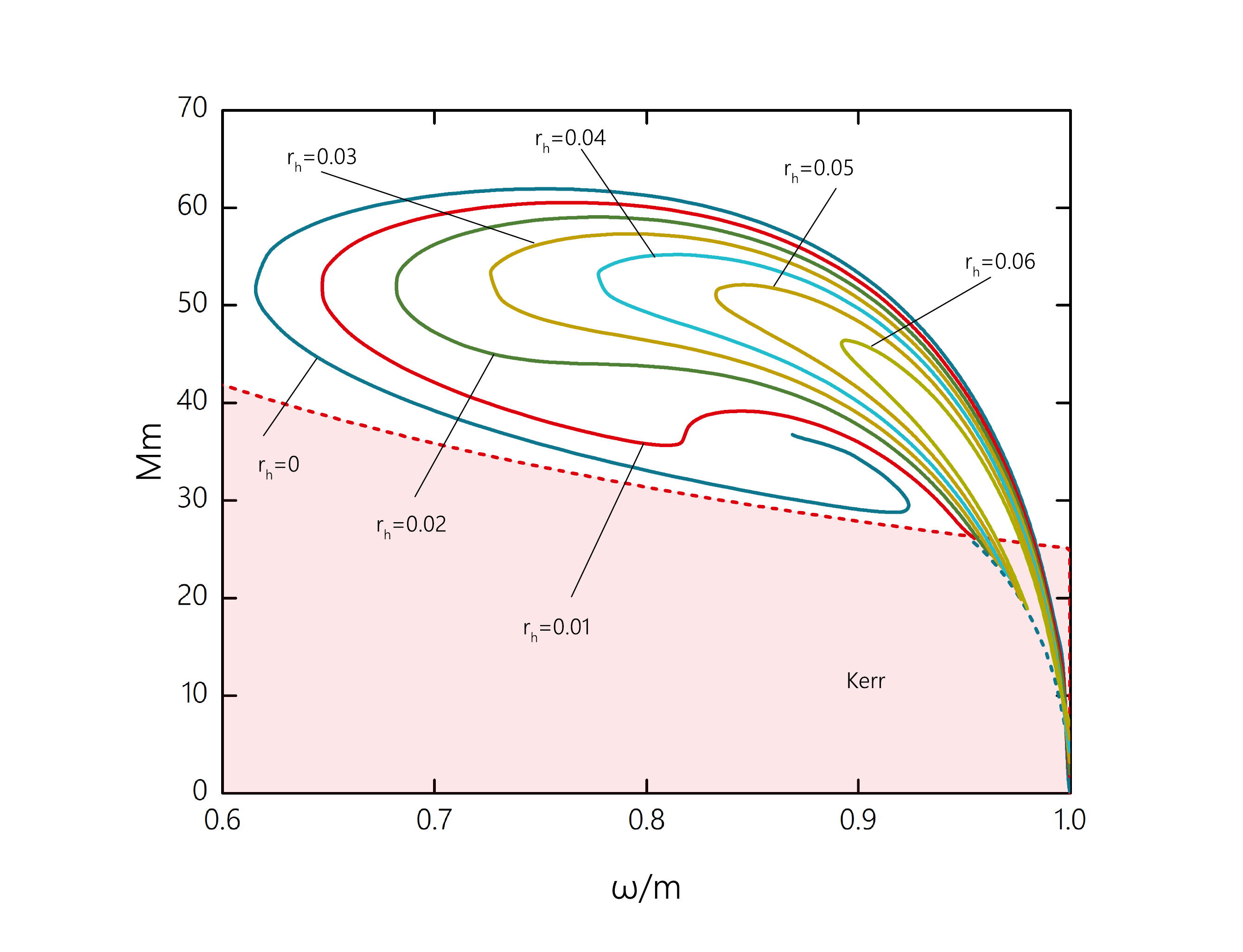}
        \includegraphics[width=.48\textwidth, trim = 40 20 90 20, clip = true]{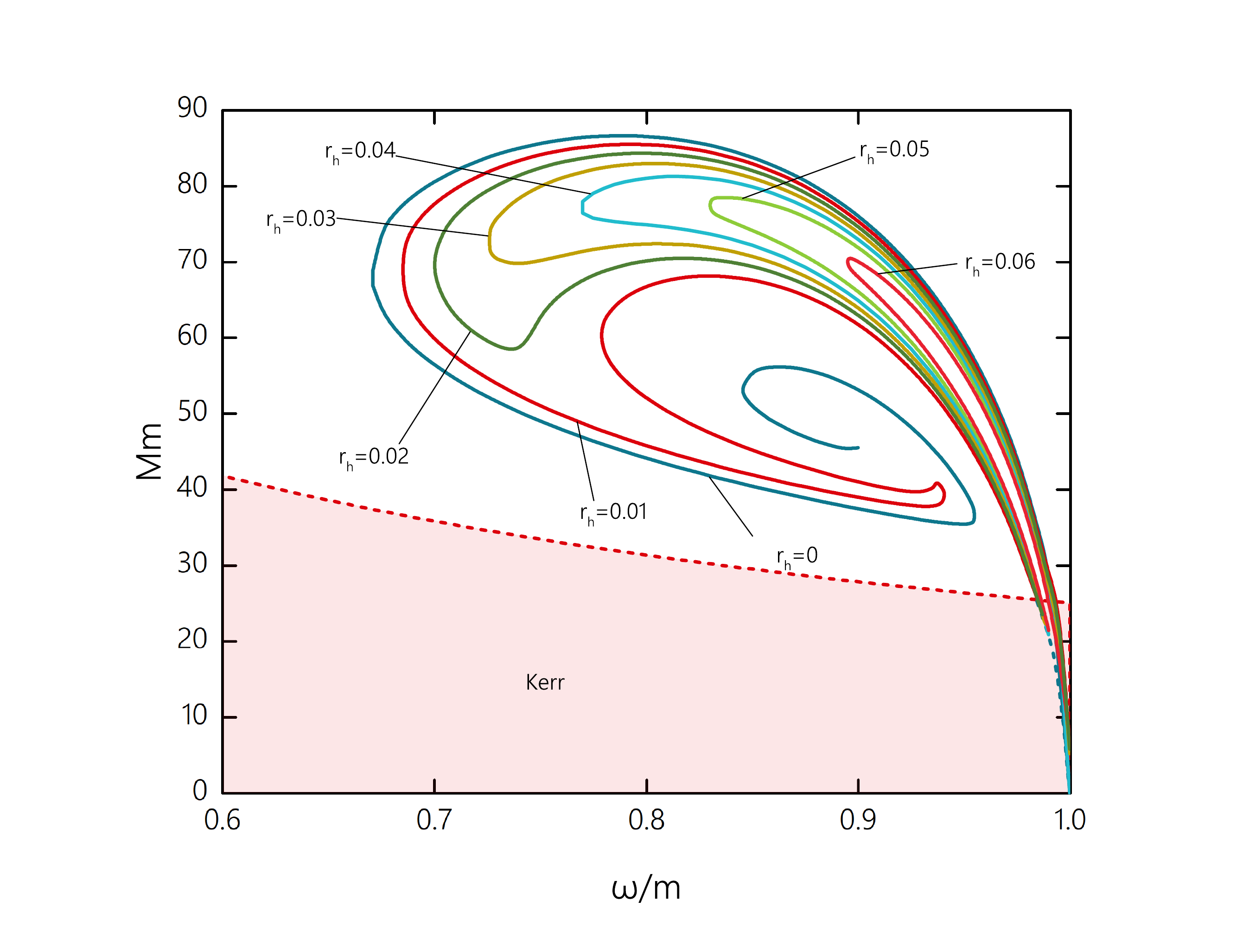}
        \includegraphics[width=.48\textwidth, trim = 40 20 90 20, clip = true]{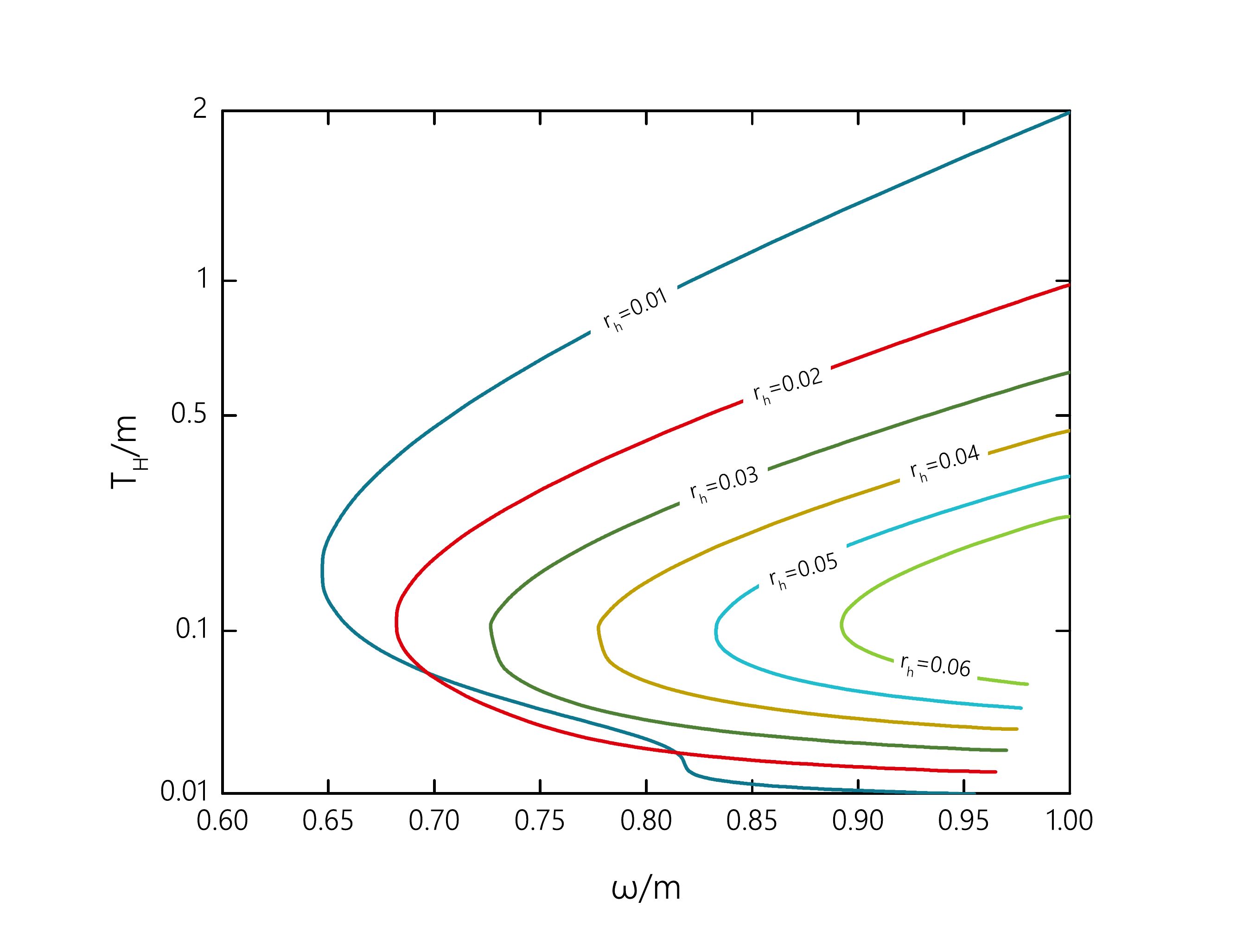}
        \includegraphics[width=.48\textwidth, trim = 40 20 90 20, clip = true]{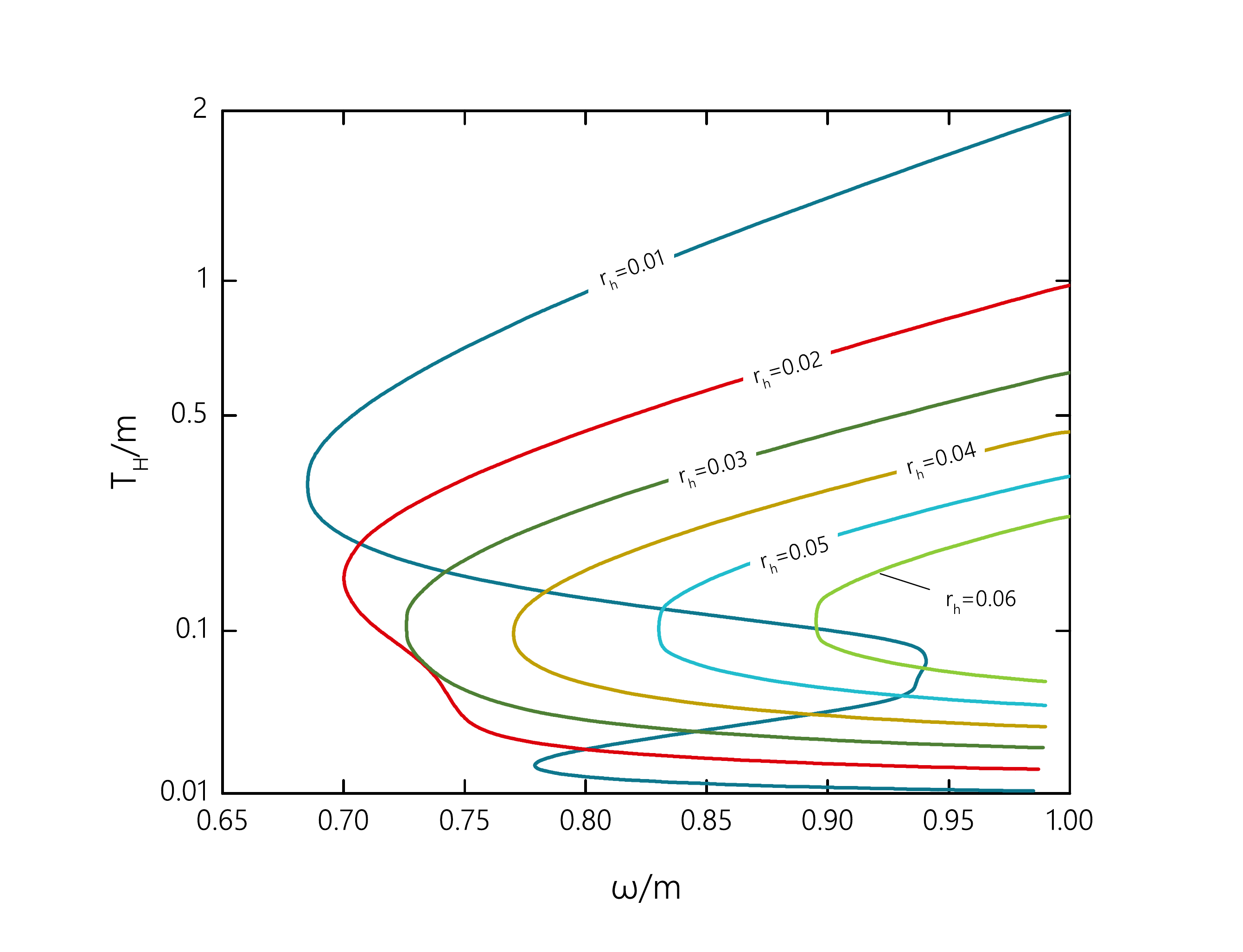}
        \includegraphics[width=.48\textwidth, trim = 40 20 90 20, clip = true]{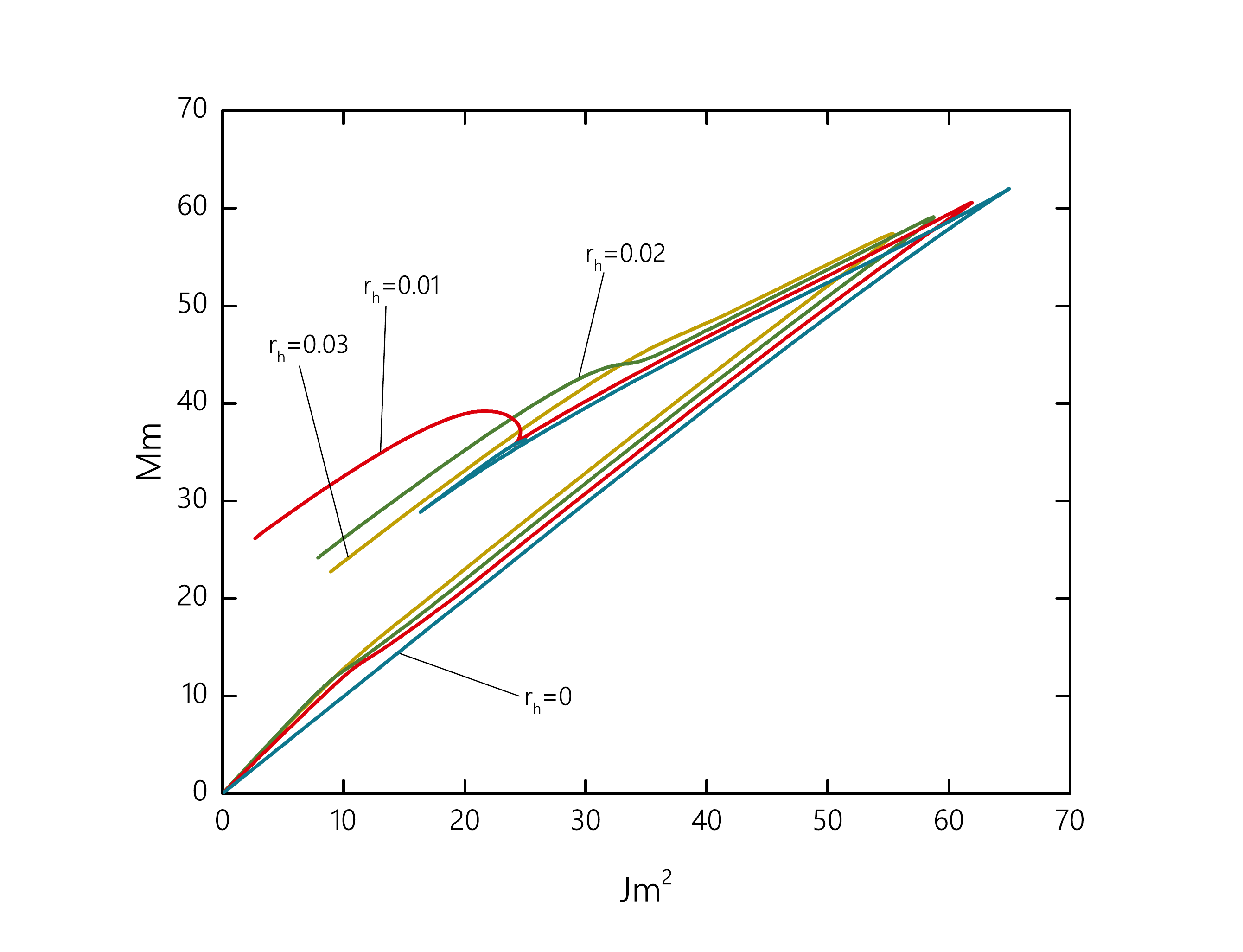}
        \includegraphics[width=.48\textwidth, trim = 40 20 90 20, clip = true]{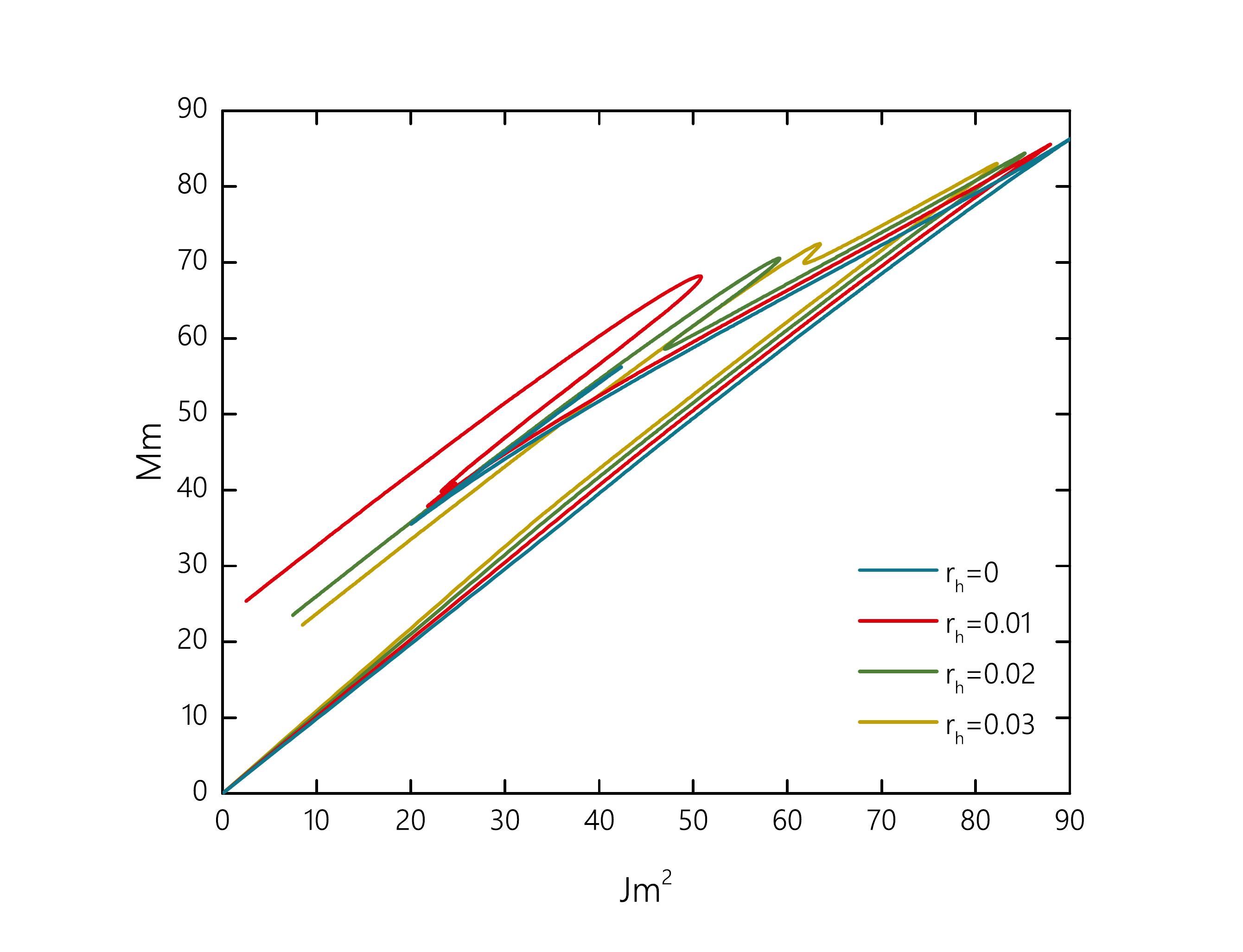}
    \end{center}
    \caption{\small
The ADM mass $M$ (upper row)
and the Hawking temperature $T_h$ (middle row)
vs the frequency $\omega$,
and the mass $M$ vs the angular momentum $J$ (bottom row)
for a set of values of the horizon radius parameter $r_h$
for $n=1$ rotating parity-even (left column) and parity-odd (right column)
hairy  black holes at $\mu/m=0.25$ and $\alpha=0.5$.
In the left plots, here, and in the subsequent figures below,
the shaded area corresponds to the domain of existence
of vacuum Kerr black holes, the red dashed line to the extremal
vacuum Kerr black holes,
and the blue dashed line to the subset of vacuum Kerr black holes
with stationary scalar Klein-Gordon clouds.}
    \lbfig{omega_rh_mu025}
\end{figure}

Note that the absence of a conical singularity on the symmetry axis
requires that the deficit angle should vanish,
$\delta=2\pi\left(1 - \lim\limits_{\theta\to 0} \frac{F_2}{F_1}\right) = 0$.
Hence the solutions should satisfy the constraint
$F_2\bigl.\bigr|_{\theta = 0}=F_1\bigl.\bigr|_{\theta = 0}$.
In our numerical scheme we explicitly
checked this condition on the symmetry axis.

\begin{figure}[hbt]
    \begin{center}
        \includegraphics[width=.48\textwidth, trim = 40 20 90 20, clip = true]{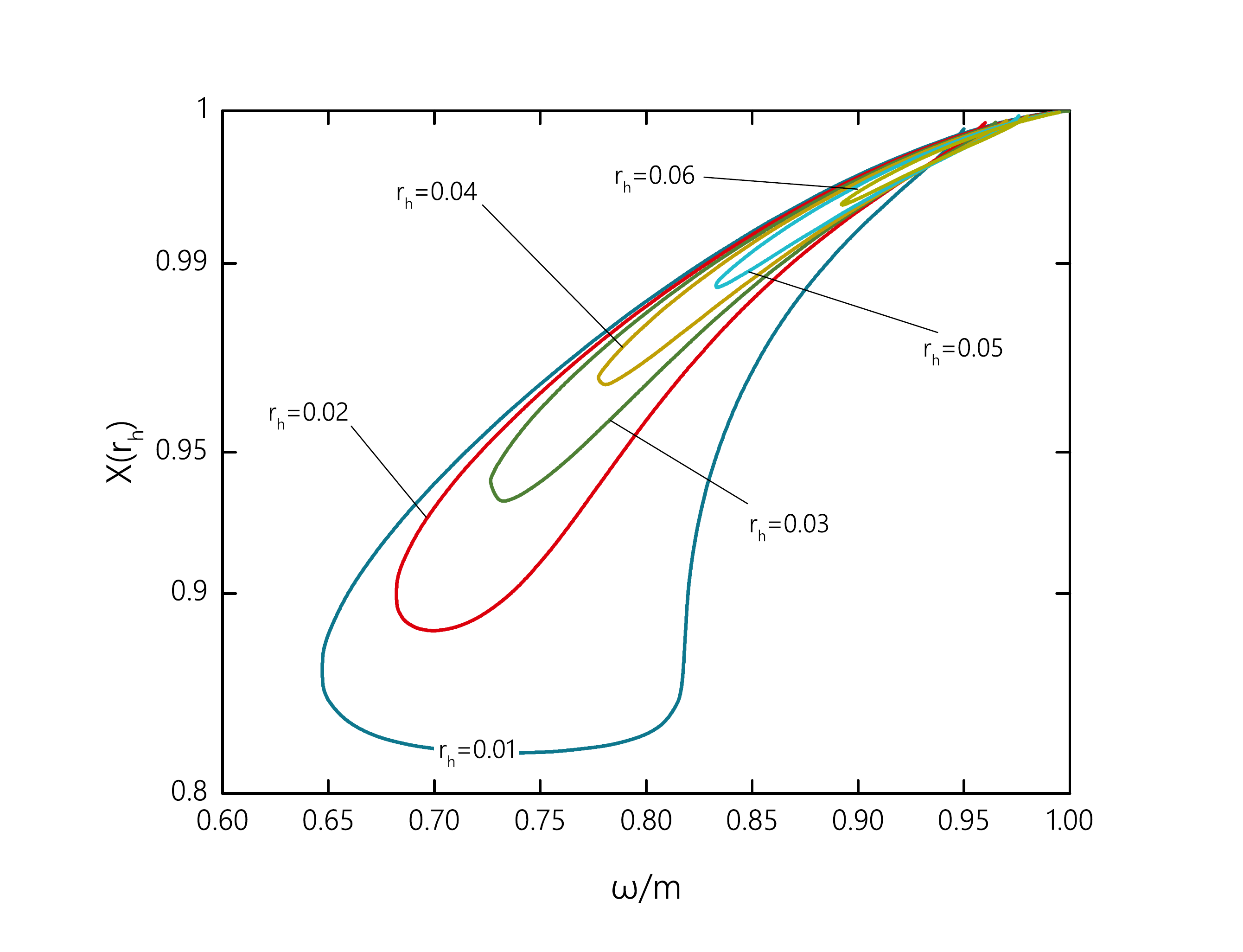}
        \includegraphics[width=.48\textwidth, trim = 40 20 90 20, clip = true]{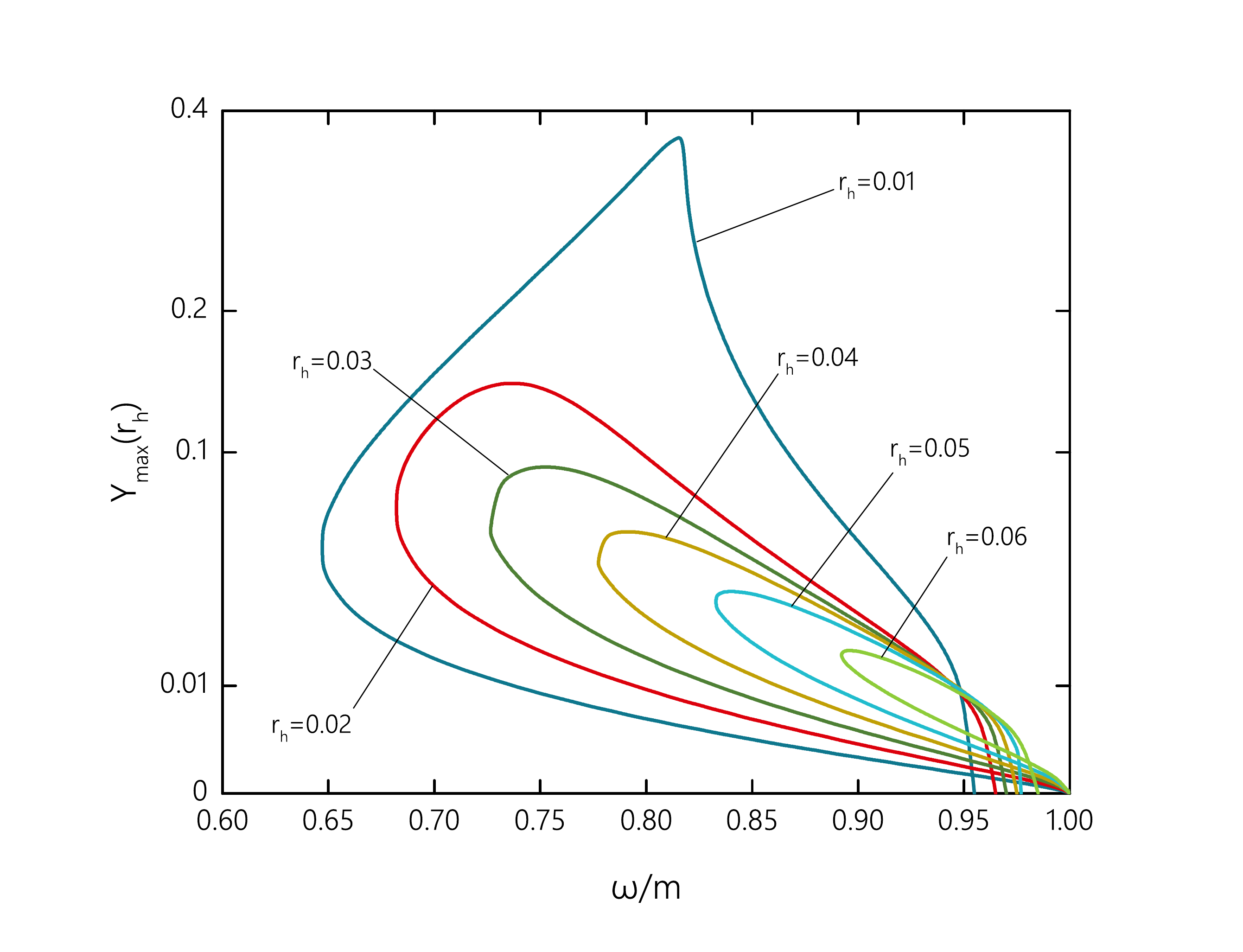}
        \includegraphics[width=.48\textwidth, trim = 40 20 90 20, clip = true]{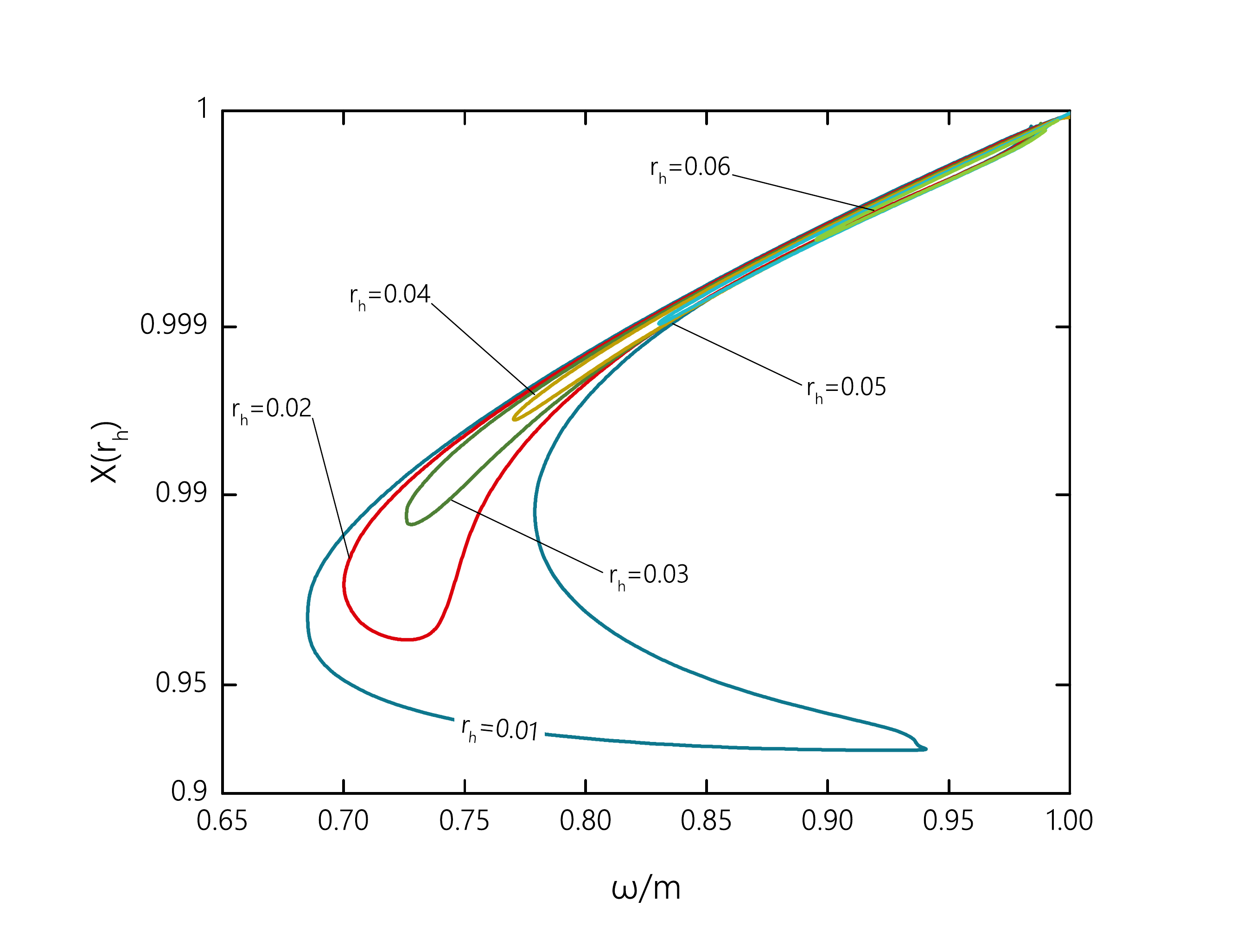}
        \includegraphics[width=.48\textwidth, trim = 40 20 90 20, clip = true]{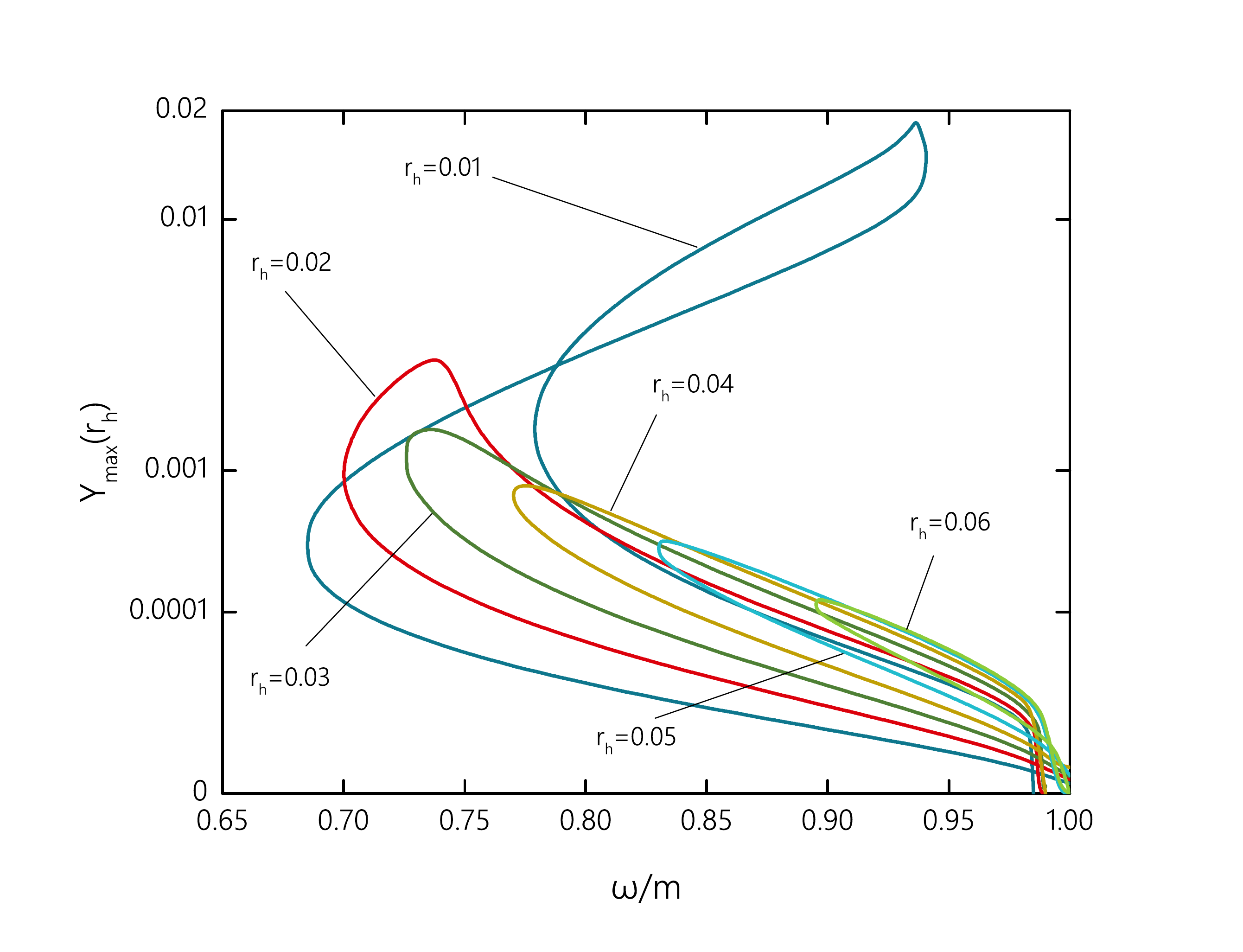}
    \end{center}
    \caption{\small
        The value of real scalar field
on the horizon $X(r_h)$ at $\theta=0$ (left column)
and the maximal value of the complex scalar field
on the horizon $Y_\mathrm{max}(r_h)$ (left column)
vs the frequency $\omega$ for a set of values of
        the horizon radius parameter $r_h$
for $n=1$ rotating parity-even (upper row) and parity-odd (lower row)
hairy black holes at $\mu/m=0.25$ and $\alpha=0.5$.}
    \lbfig{fields_rh_omega_rh_mu025}
\end{figure}

\begin{figure}[hbt]
    \begin{center}
        \includegraphics[width=.8\textwidth, trim = 40 20 90 20, clip = true]{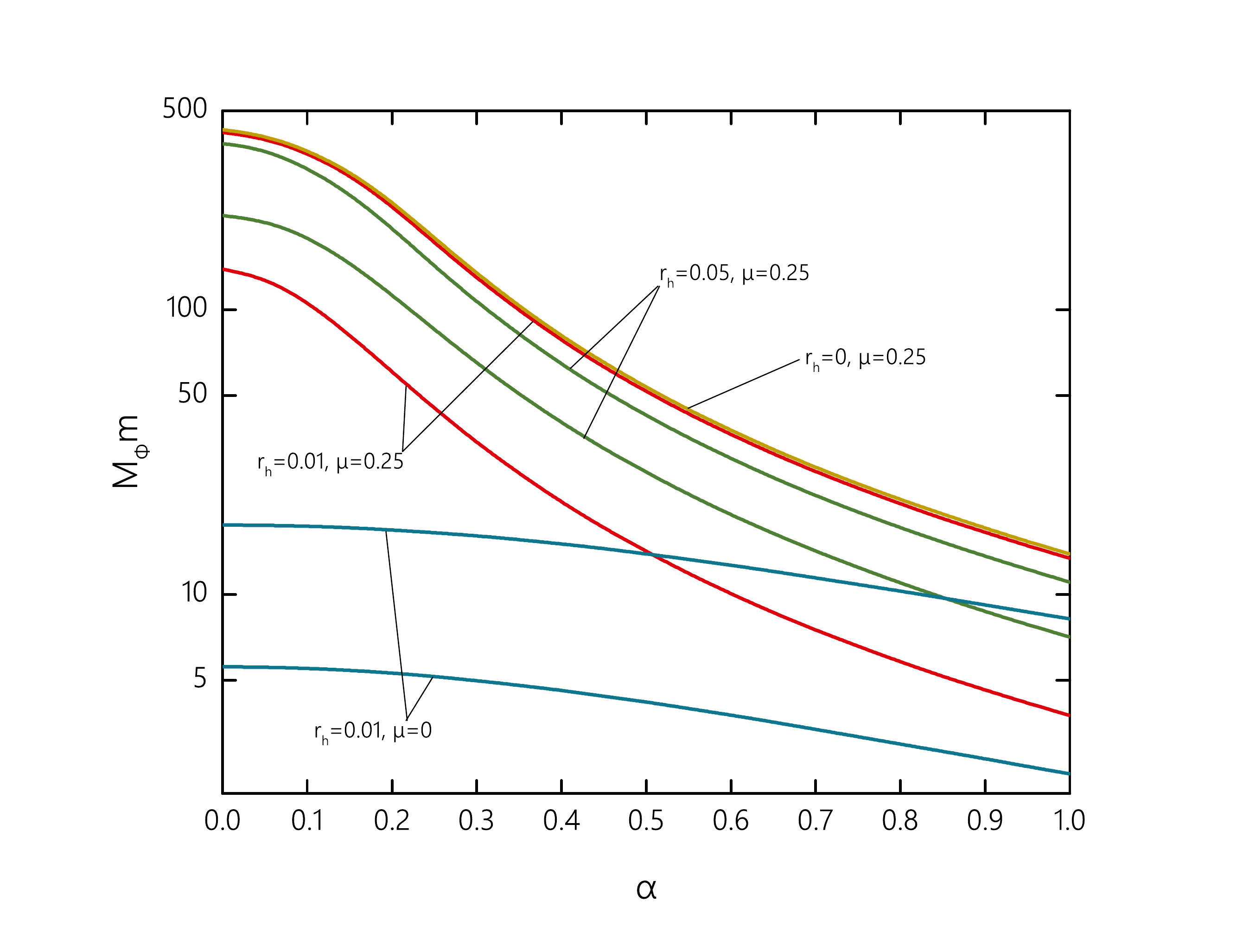}
    \end{center}
    \caption{\small
        The scalar hair mass $M_\Phi$ vs the gravitational coupling constant
        $\alpha$ for a set of values of horizon radius parameter $r_h$ and
        the potential coupling constant $\mu$
for $n=1$ rotating parity-even hairy black holes at frequency $\omega/m=0.9$.
        }
    \lbfig{alpha}
\end{figure}

\subsection{Quantities of interest and Smarr relation}

Asymptotic expansions of the metric functions at the horizon
and at spatial infinity yield a number of physical observables.
The total ADM mass $M$ and the angular momentum $J$
of the spinning hairy black holes can be read off
from the asymptotic subleading behaviour
of the metric functions as $r\to \infty$
\be
\label{ADM}
g_{tt}=-1 + \frac{\alpha^2 M}{\pi r}+O\left(\frac{1}{r^2}\right), \quad
g_{\varphi t}=\frac{\alpha^2 J}{\pi r}\sin^2 \theta+O\left(\frac{1}{r^2}\right).
\ee
\begin{figure}[hbt]
    \begin{center}
        \includegraphics[width=.24\textwidth, clip = true]{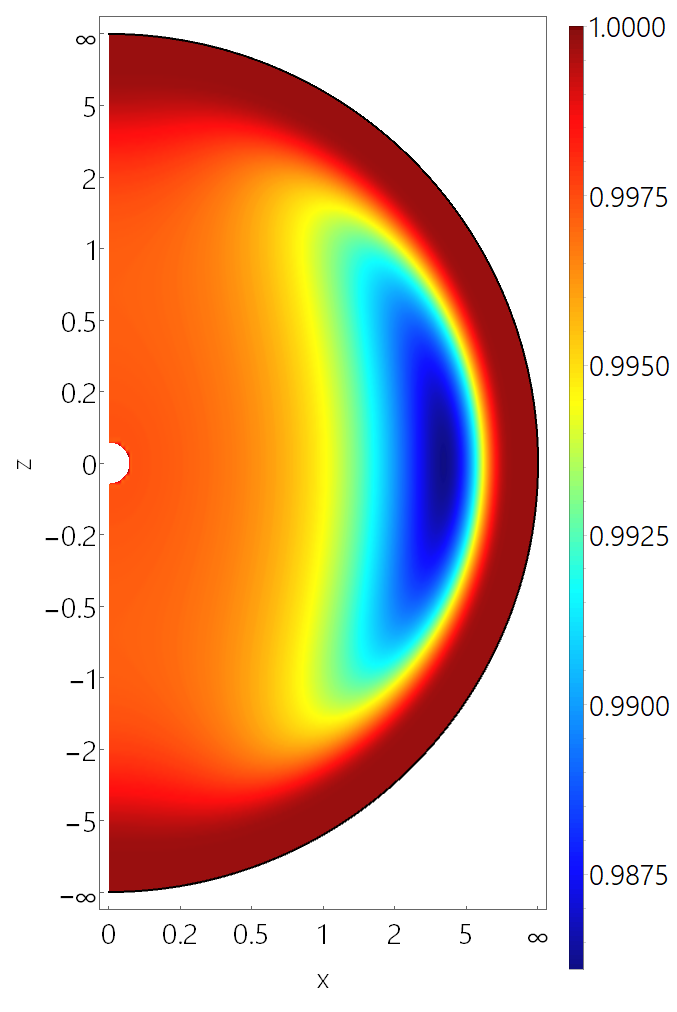}
        \includegraphics[width=.24\textwidth, clip = true]{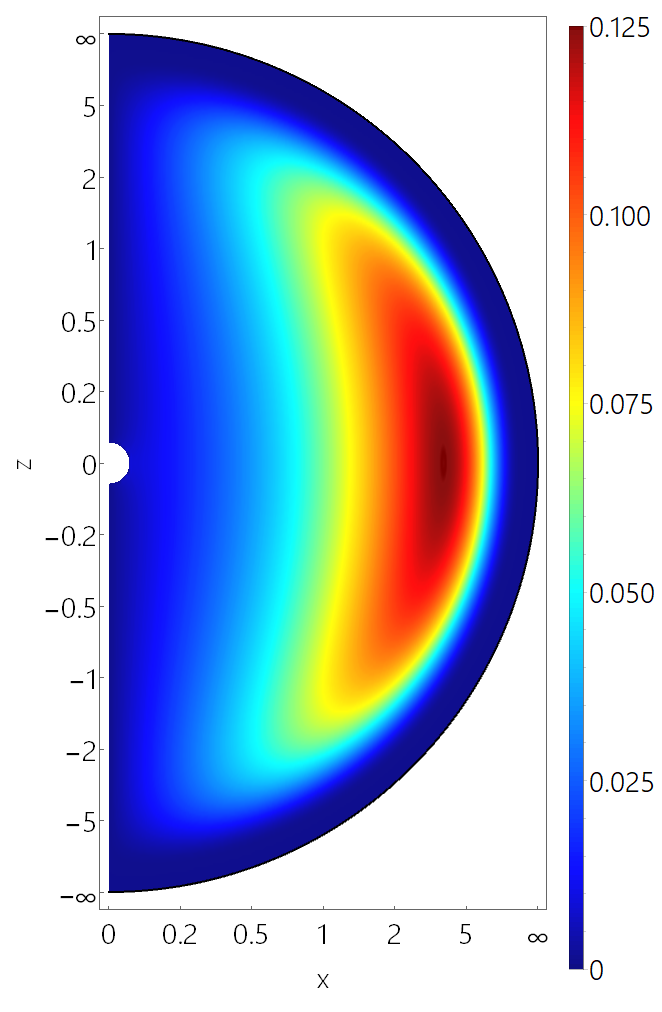}
        \includegraphics[width=.24\textwidth, clip = true]{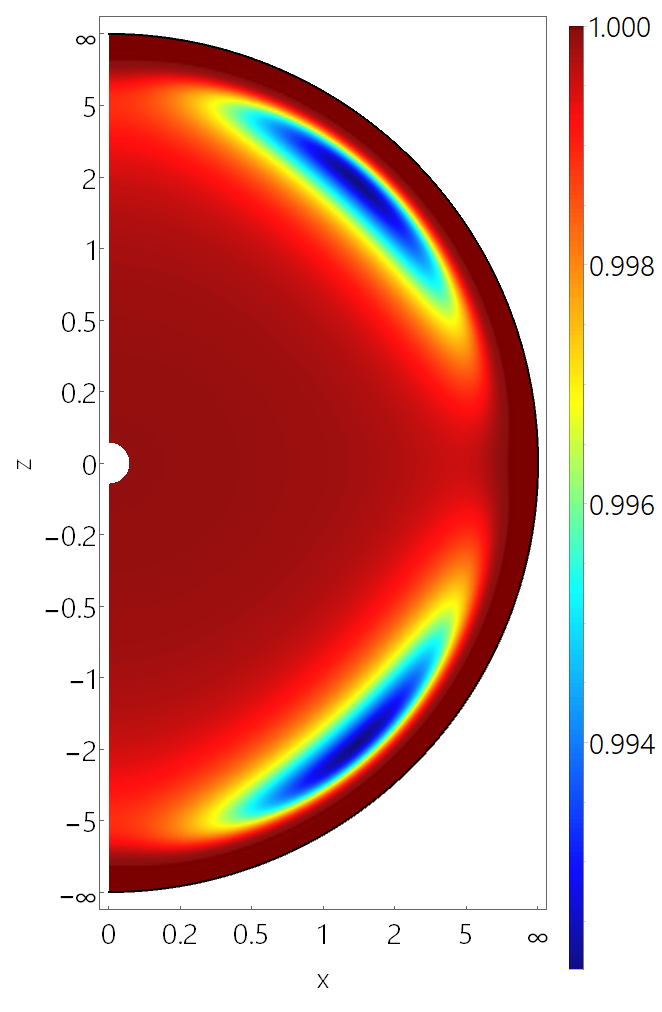}
        \includegraphics[width=.24\textwidth, clip = true]{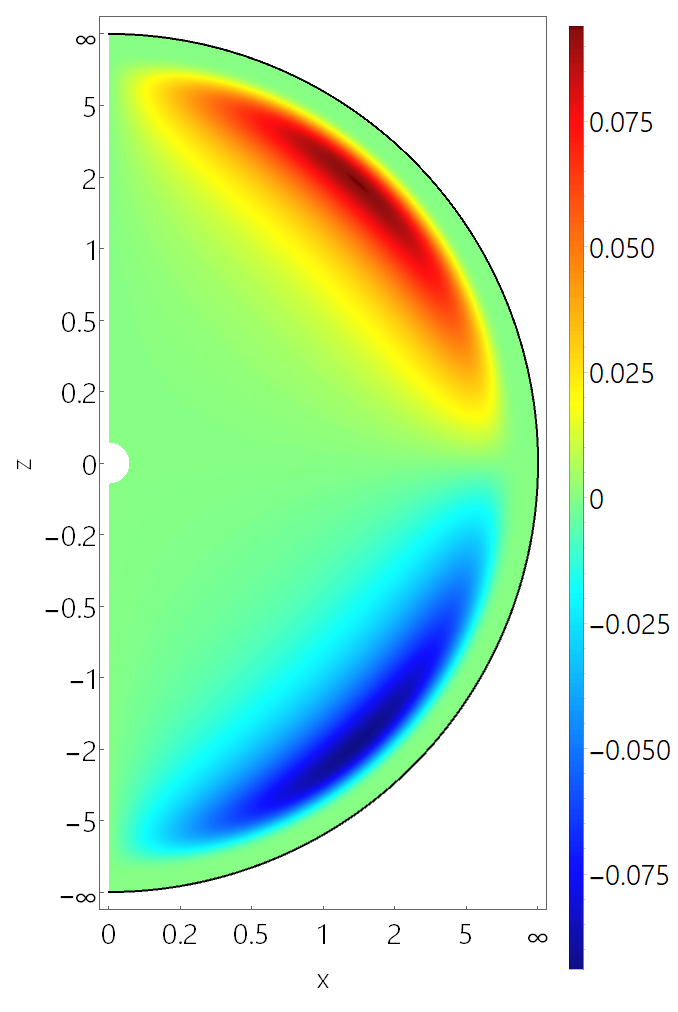}
        \includegraphics[width=.24\textwidth, clip = true]{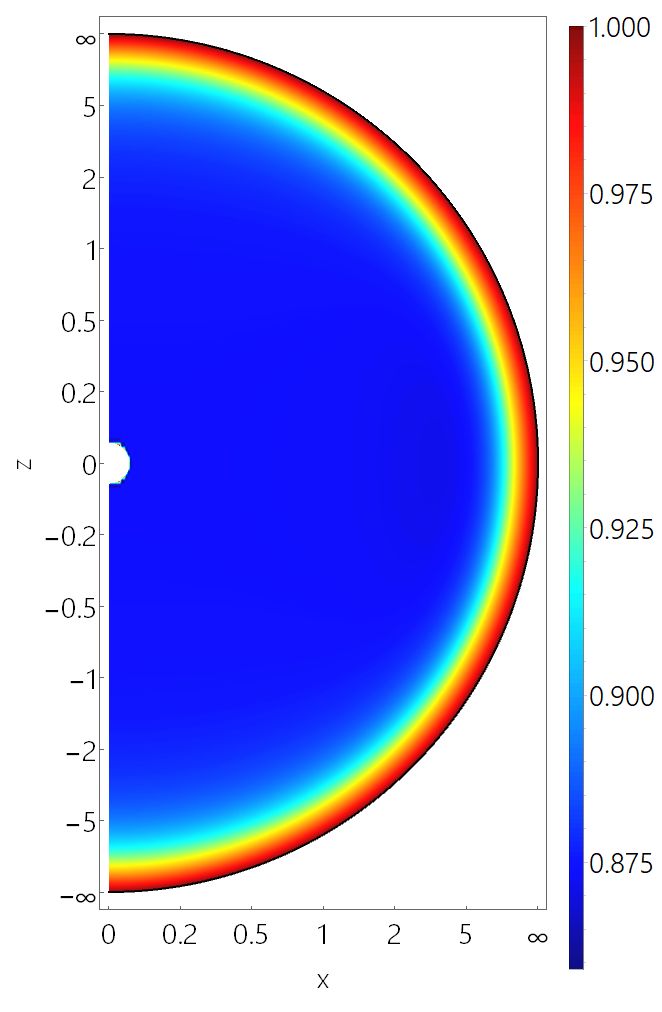}
        \includegraphics[width=.24\textwidth, clip = true]{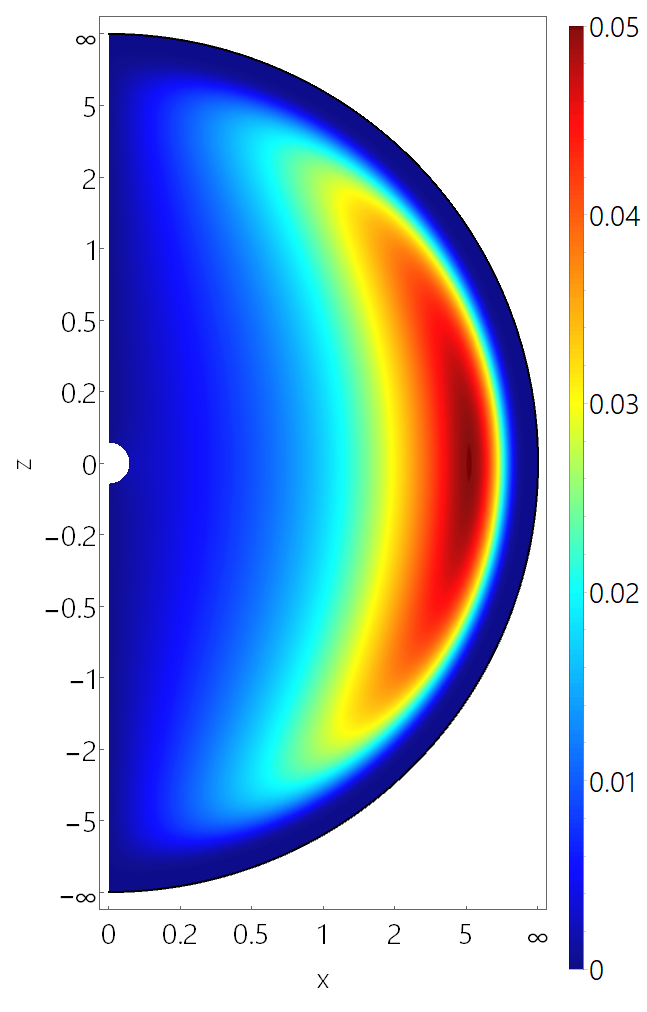}
        \includegraphics[width=.24\textwidth, clip = true]{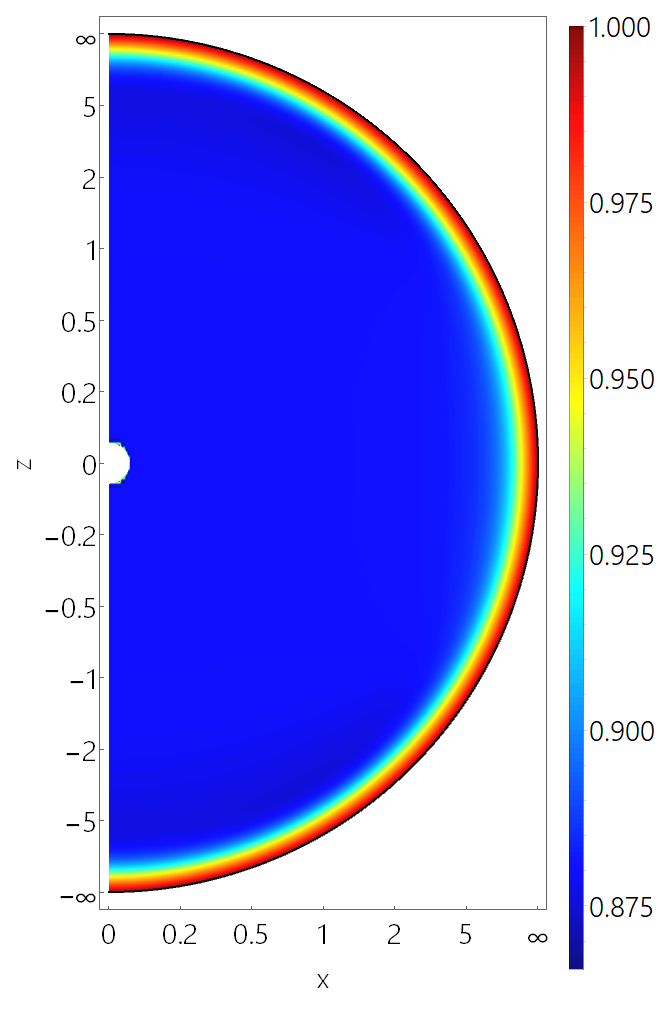}
        \includegraphics[width=.24\textwidth, clip = true]{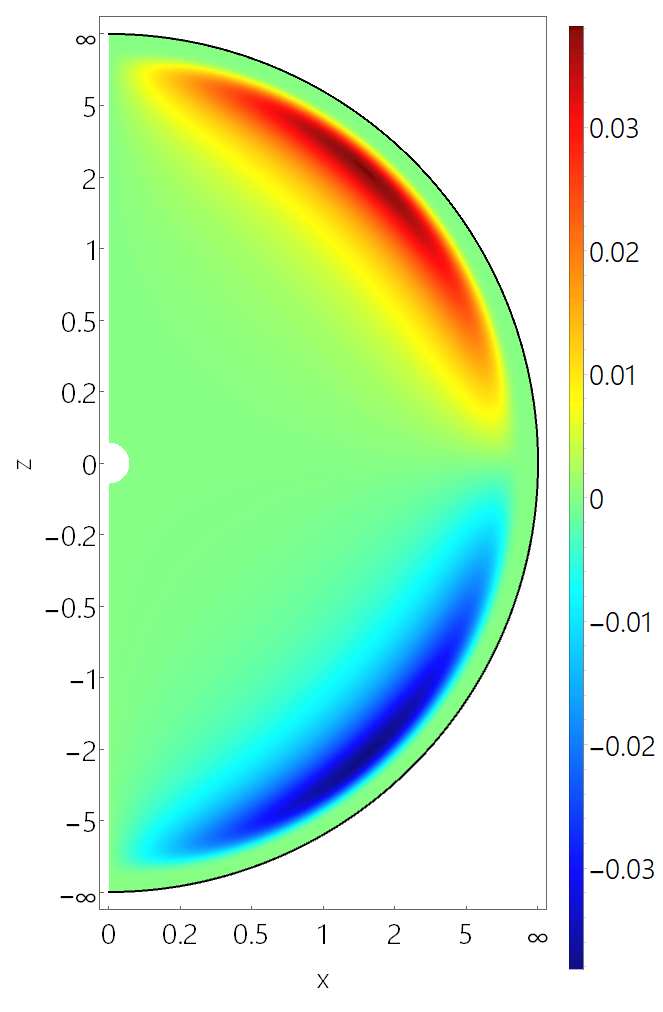}
    \end{center}
    \caption{\small
        Contour plots  of the profile function of the real scalar field $X$
(first and third columns) and
        the profile function of the complex scalar field $Y$
(second and fourth columns) in the $y=0$ plane for
        parity-even (left two columns)
and parity-odd (right two columns)
$n=1$ black holes with synchronized hair
        at $\omega/m=0.9, \alpha=0.5$ and horizon radius parameter $r_h=0.05$.
        The upper row shows solutions for a finite value of
the real scalar field mass $\mu/m=0.25$,
whilst the bottom row shows solutions for a massless real scalar field with
        $\mu=0$.}
    \lbfig{fields-profiles}
\end{figure}

The ADM charges can be represented as sums of the contributions
from the event horizon and the scalar hair,
$M = M_h+M_\Phi$ and $J = J_h+J_\Phi$, respectively.
These contributions can be evaluated separately via Komar integrals
\be
\label{Komar}
\begin{split}
    M_h&=-\frac{1}{2\alpha^2}\oint_\mathrm{S}{dS_{\mu\nu} \nabla^\mu \xi^\nu},
    \quad J_h=\frac{1}{4\alpha^2}\oint_\mathrm{S}{dS_{\mu\nu} \nabla^\mu \eta^\nu},\\
    M_\Phi&=-\frac{1}{\alpha^2}\int_{V}{dS_{\mu} \left(2 T^\mu_\nu \xi^\nu - T\xi^\mu\right)},
    \quad J_\Phi=\frac{1}{2\alpha^2}\int_\mathrm{V}{dS_{\mu} \left(T^\mu_\nu \eta^\nu - \frac12 T\eta^\mu\right)},
\end{split}
\ee
where $S$ is the horizon 2-sphere and $V$ denotes an asymptotically flat
spacelike hypersurface bounded by the horizon.

\begin{figure}[hbt]
    \begin{center}
        \includegraphics[width=.48\textwidth, trim = 40 20 90 20, clip = true]{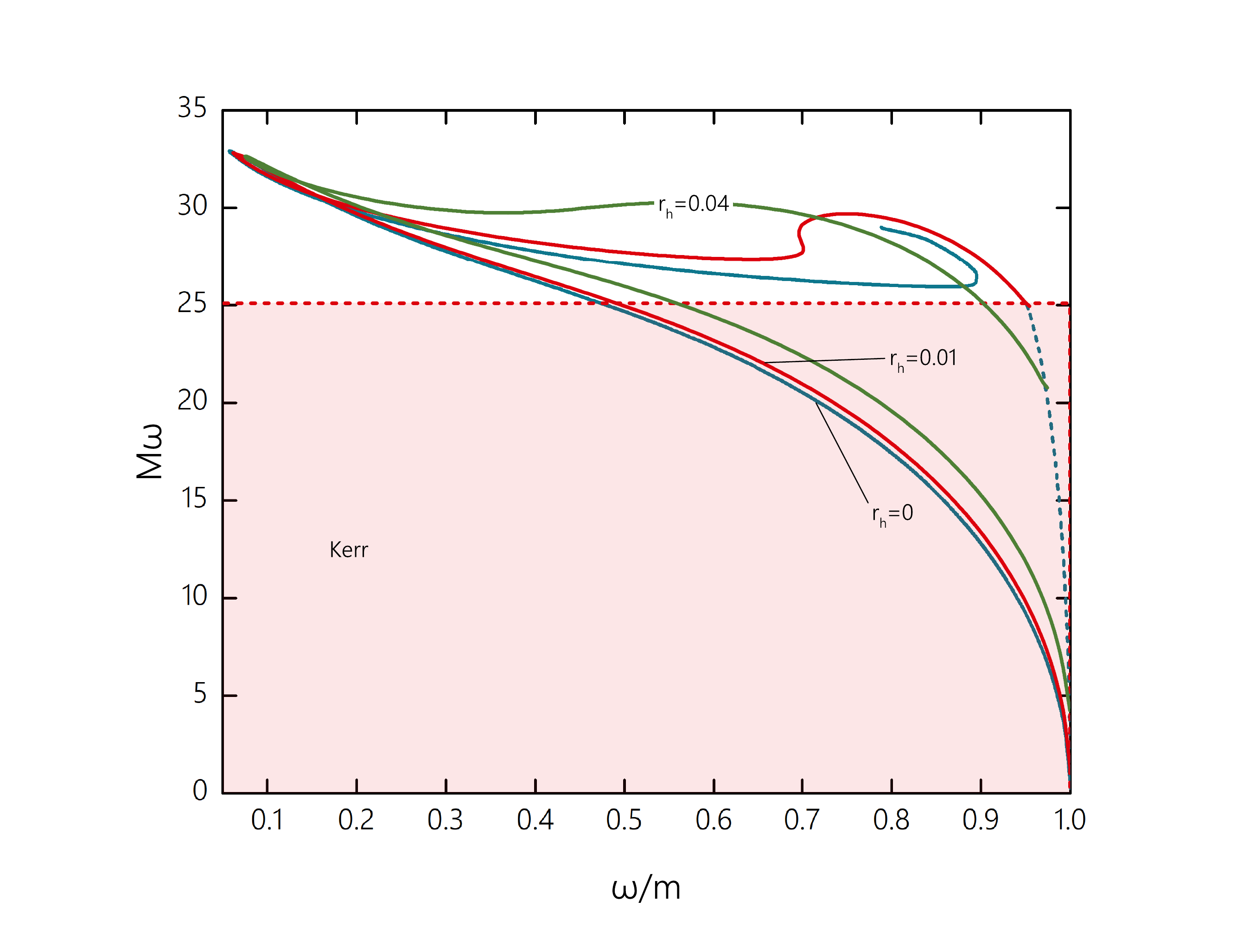}
        \includegraphics[width=.48\textwidth, trim = 40 20 90 20, clip = true]{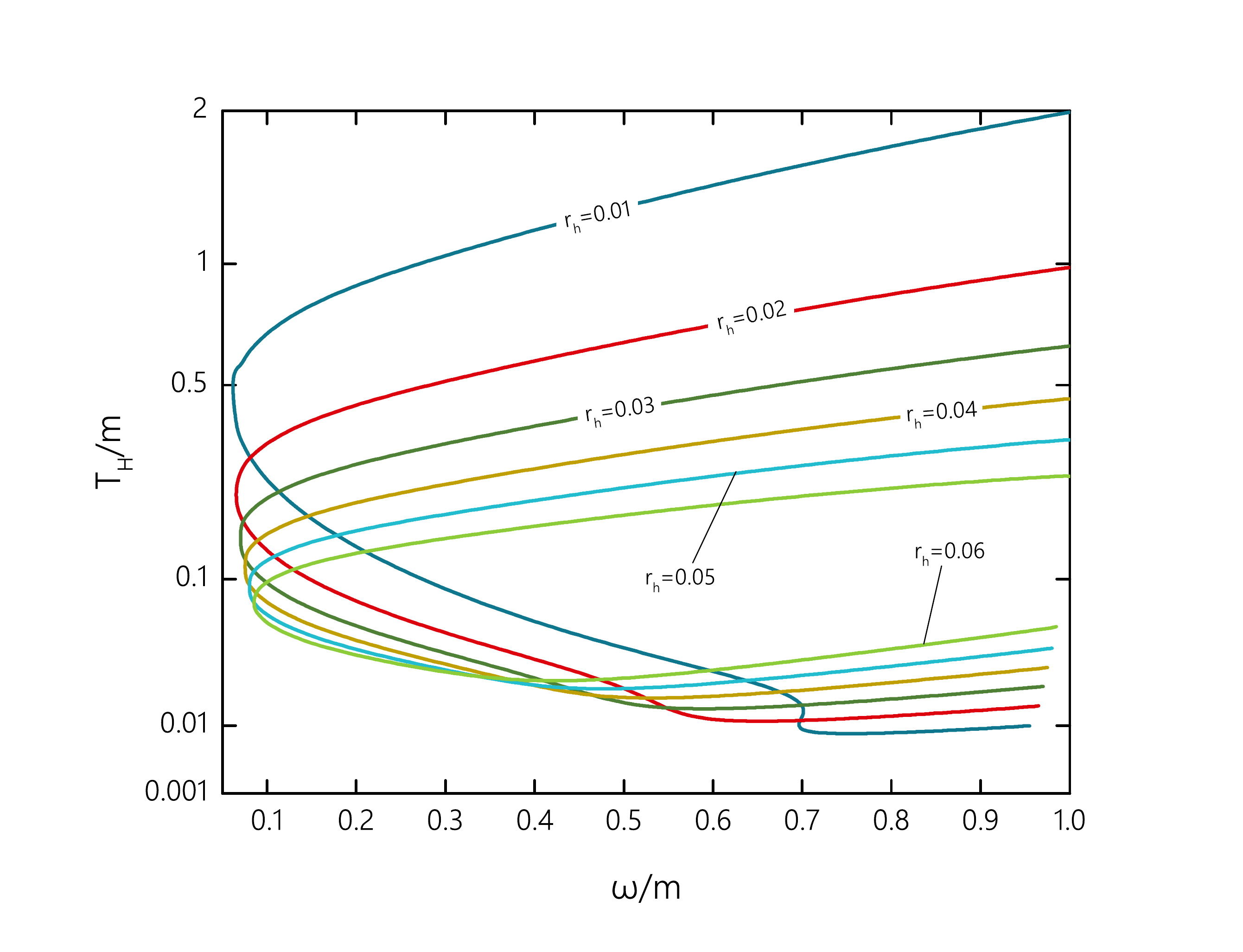}
    \end{center}
    \caption{\small
        The ADM mass $M$ (left plot)
and the Hawking temperature $T_h$ (right plot)
        vs the frequency $\omega$ for a set of values of
        the horizon radius parameter $r_h$
for $n=1$ rotating parity-even hairy black holes
 in the limiting case of vanishing scalar potential ($\mu=0$) at $\alpha=0.5$.
 To simplify the presentation the ADM mass is multiplied by $\omega$.}
    \lbfig{omega_rh_mu0}
\end{figure}

Analogously to other axially symmetric stationary rotating boson stars
\cite{Kleihaus:2005me,Kleihaus:2007vk},
self-gravitating solitons of the non-linear sigma model \cite{Herdeiro:2018djx} and vortons \cite{Kunz:2013wka},
one obtains the quantization relation for the angular momentum
of the spinning complex scalar component of the model,
$J_\Phi=n Q$, where $Q$ is its Noether charge
and $n$ its winding number.
\begin{figure}[hbt]
\begin{center}
    \includegraphics[width=1\textwidth, trim = 20 20 20 20, clip = true]{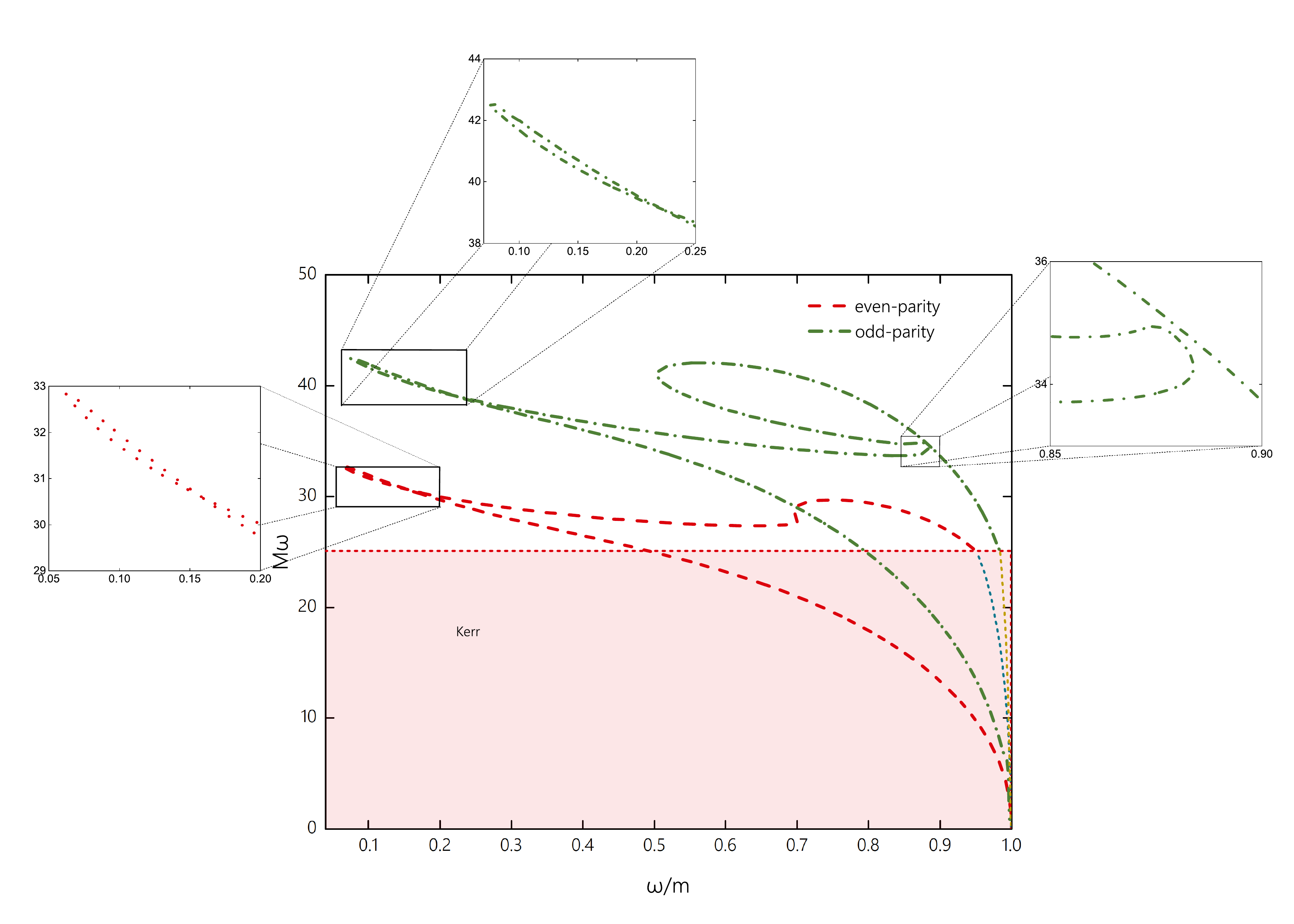}
\end{center}
\caption{\small
    The ADM mass $M$ vs the frequency $\omega$ for
parity-even and parity-odd
    $n=1$  rotating hairy black holes at $\mu=0, r_h=0.01$ and $\alpha=0.5$.
To simplify the presentation
    the ADM mass is multiplied by $\omega$.}
\lbfig{parity}
\end{figure}

\begin{figure}[hbt]
    \begin{center}
        \includegraphics[width=.48\textwidth, trim = 40 20 90 20, clip = true]{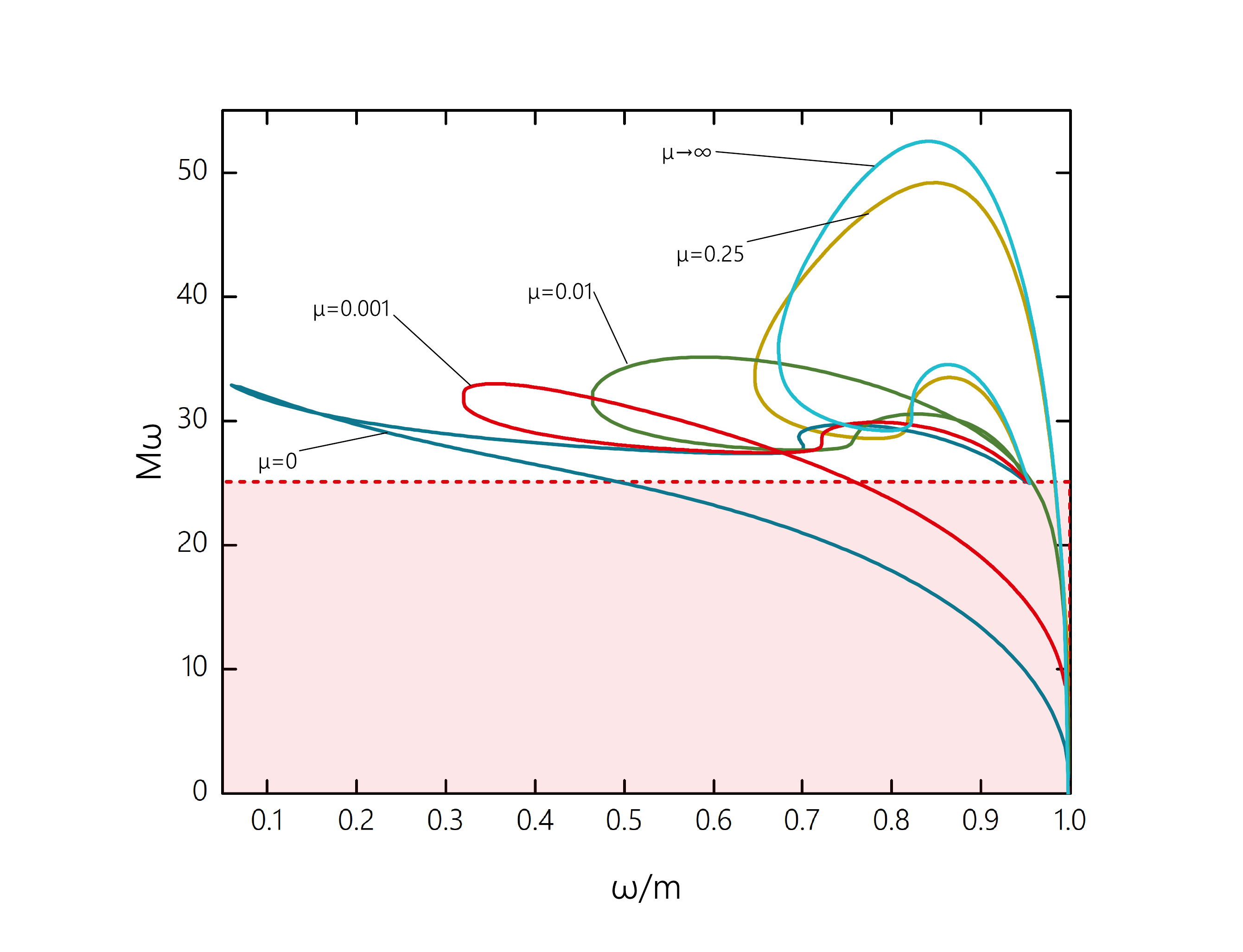}
        \includegraphics[width=.48\textwidth, trim = 40 20 90 20, clip = true]{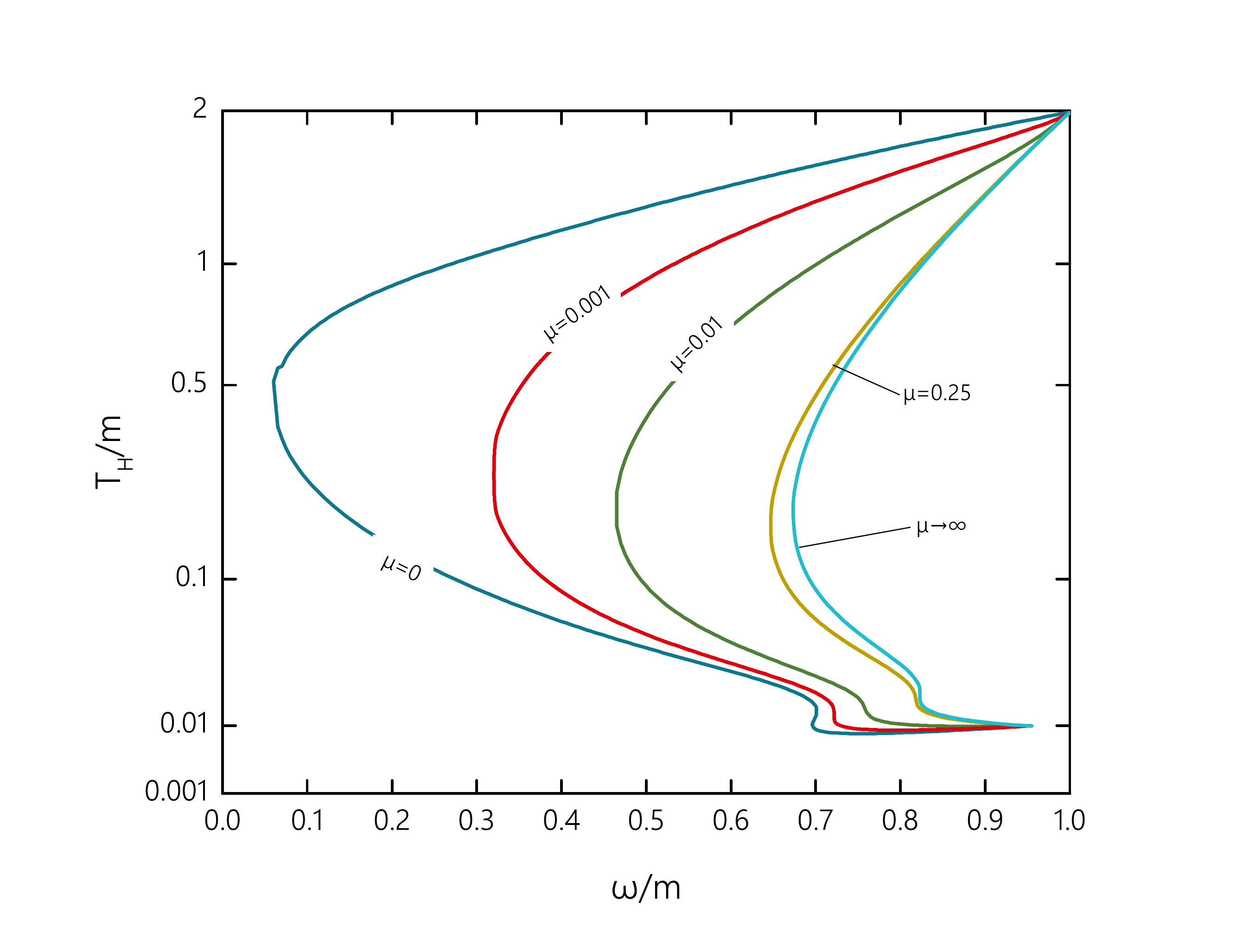}
        \includegraphics[width=.48\textwidth, trim = 40 20 90 20, clip = true]{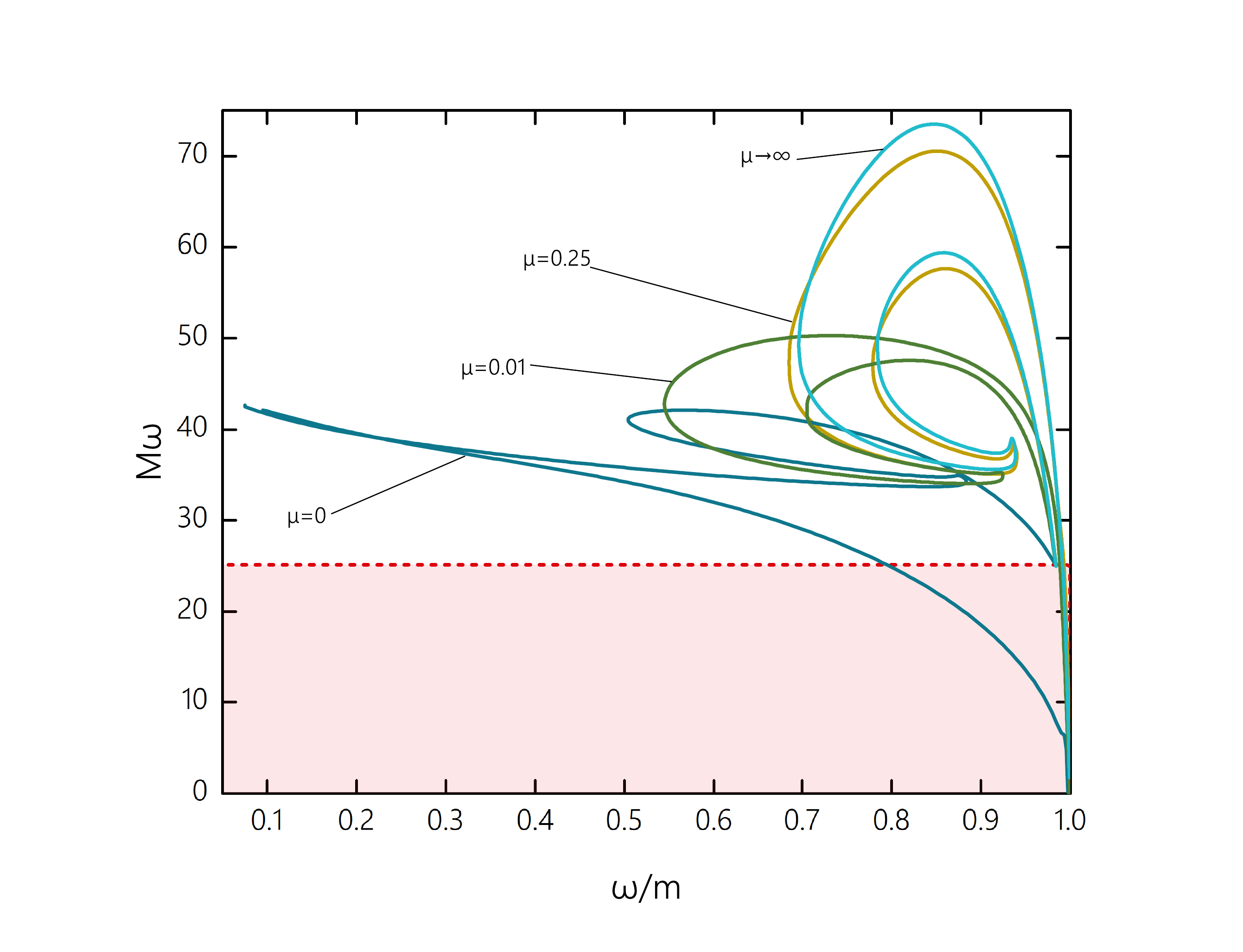}
        \includegraphics[width=.48\textwidth, trim = 40 20 90 20, clip = true]{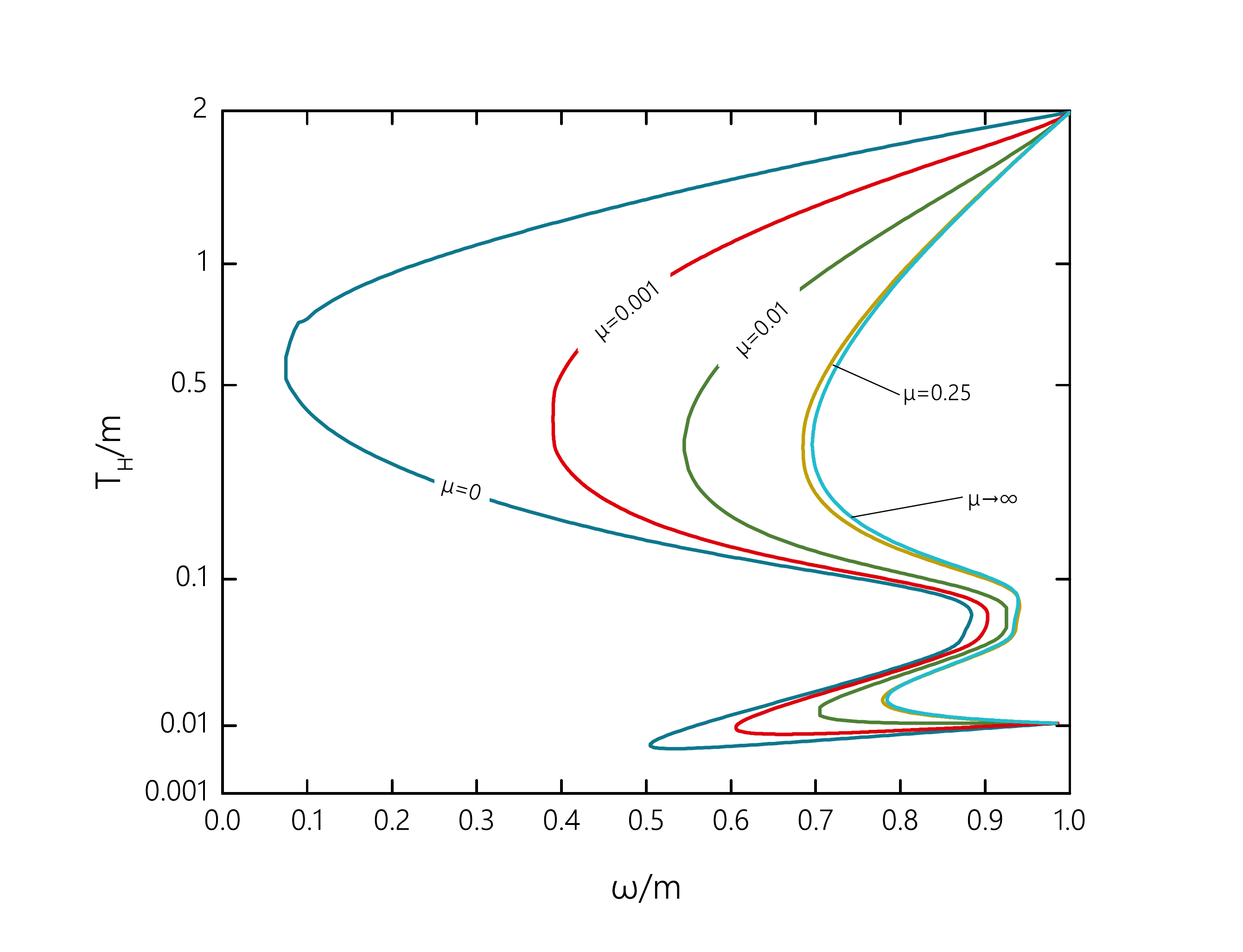}
    \end{center}
    \caption{\small
        The ADM mass $M$ (left column)
and the Hawking temperature $T_h$ (right column)
        vs the frequency $\omega$ for a set of values of
        the potential coupling constant $\mu/m$
for $n=1$ parity-even (upper row) and parity-odd (lower row)
rotating hairy black holes at the horizon radius parameter $r_h=0.01$ and $\alpha=0.5$.}
    \lbfig{omega_mu}
\end{figure}

The physically interesting horizon properties include the Hawking
temperature $T_h$, which is proportional to the surface gravity
$\kappa^2=-\frac{1}{2}\nabla_\mu\chi_\nu\nabla^\mu\chi^\nu$
\be
\label{horQ}
T_h=\frac{\kappa}{2\pi} = \frac{1}{16\pi r_h}\exp\left[\left(f_0 - f_1\right)\bigl.\bigr|_{r = r_h}\right]\, ,
\ee
where $\chi$ is the horizon Killing vector \re{Killingrh}.
Of interest is also the horizon area $A_h$, which is given by
\be
\label{horQ-A}
A_h = 32\pi r_h^2 \int_0^\pi d\theta \sin\theta\exp\left[\left(f_1 + f_2\right)\bigl.\bigr|_{r = r_h}\right] \, .
\ee
The observables are related via the Smarr relation
\be
\label{Smarr}
M = 2T_h S + 2\Omega_h J_h + M_\Phi \, ,
\ee
where $S=\frac{\pi}{\alpha^2}A_h$ is the entropy of the black hole
and $M_\Phi$ is the energy of the scalar fields outside
the event horizon \re{Komar}.
Another relation between the physical quantities
of the hairy black hole is the first law of thermodynamics
$$
dM = T_h dS + \Omega_h d J \, .
$$

\section{Results}
\subsection{Numerical implementation}
The set of six coupled non-linear elliptic partial differential equations
of the functions $X,Y,W,f_0,f_1,f_2$, which
parametrize the system \re{Einstein},\re{scaleq},
has been solved numerically subject to the boundary conditions
\re{bchor}-\re{bcpole}.
We have made use of a fourth-order finite differences scheme,
where the system of equations is discretized on a grid
with $101\times 101$ points.
To facilitate the calculations in the near horizon area,
we have introduced the new radial coordinate $x=\frac{r-r_h}{r+c}$,
which maps the semi-infinite region $[0,\infty )$
onto the unit interval $[0,1]$.
Here $c$ is an arbitrary constant used to adjust the contraction of the grid.
The emerging system of nonlinear algebraic equations is solved
using a modified Newton method.
The underlying linear system is solved with the
Intel MKL PARDISO sparse direct solver.
The errors are on the order of $10^{-5}$.
All calculation have been performed using CESDSOL
\footnote{Complex Equations -- Simple Domain partial differential equations
SOLver is a C++ library developed by IP.
It provides tools for the discretization of an arbitrary number
of arbitrarily nonlinear equations with arbitrary boundary conditions
on direct product arbitrary dimensional grids
with arbitrary order of accuracy.} library.

\subsection{Stationary spinning self-gravitating solitons and black holes
with synchronised hair}
When constructing spinning gravitating solutions,
we start with the corresponding flat spacetime
configurations \cite{Loiko:2018mhb} as initial guesses.
Since the present system possesses a large space
of parameters, we restrict our analysis to two particular values
of the mass parameter $\mu$, $\mu^2=0.25$ and
$\mu=0$ (i.e., the limiting case of vanishing potential),
recall that the mass $m$ of the complex field can be rescaled to $m=1$.
The regular self-gravitating solitonic solutions are found by
replacing the boundary conditions at the
horizon $r_h$ with appropriate boundary conditions at $r=0$.

As usual, the stationary spinning solutions exist within
a restricted interval of the scaled frequency
$\omega\in\left[\omega_\mathrm{min},\omega_\mathrm{max}=1\right]$.
The upper bound of the frequency $\omega$
corresponds to the mass of the complex scalar field,
the lower critical frequency depends on the
value of the second mass parameter $\mu$.
As $\omega \to \omega_\mathrm{max}=1$
the spinning configurations smoothly approach linearized
perturbations around Minkowski spacetime,
with the ADM mass $M$ tending to zero.

Note that, unlike the corresponding solutions
of the non-renormalizable flat space model with a single complex field
and a sextic potential \cite{Kleihaus:2005me,Kleihaus:2007vk},
there is no lower bound on the frequency in the flat space limit
of the present model:
the Friedberg-Lee-Sirlin Q-balls exist for all non-zero values
of the frequency $\omega<m$ \cite{Levin:2010gp,Loiko:2018mhb}.
As we shall see,
the coupling to gravity changes this,
and a lower critical value of the frequency appears.

Setting the winding number $n$ to $n=0$
reduces the above system to the spherically symmetric case
of the regular boson star solutions. In Fig.~\ref{M_n0}
we display the ADM mass $M$  as a function of rescaled frequency (left) and the  mass $M$
vs the charge $Q$ for oscillating $n=0$ parity-even boson stars in the model \re{action}.

Notably, the coupling to gravity reduces the interval of values of the frequency $\omega$,
and the spherically symmetric boson stars exist in
a more limited frequency range. The minimal value $\omega_{min}>0$ depends on the strength of the gravitational coupling $\alpha$ and on the value of the mass parameter $\mu$, see Fig.~\ref{M_n0}, left plot.
As expected, these boson stars exhibit a typical spiral-like
frequency dependence of the charge and the mass,
and should approach finite limiting values at the centers of the spirals.

Considering the $n=0$ configurations we do not find spherically symmetric hairy black holes, as expected.
However, taking $n \ge 1$, we obtain both boson stars
with non-zero angular momentum, and spinning hairy black holes.
Furthermore, for $n\ge 1$ there are two different types
of spinning configurations,
\cite{Volkov:2002aj,Kleihaus:2005me,Kleihaus:2007vk,Wang:2018xhw,Kunz:2019bhm},
referred to as parity-even and
parity-odd solutions, respectively.

In Fig.~\ref{omega_rh_mu025} the ADM mass $M$ of the $n=1$ parity-even
and parity-odd solutions is exhibited versus the frequency $\omega$
for a set of values of the horizon radius parameter $r_h$,
including the regular limit $r_h=0$,
for the chosen values of the mass parameter $\mu^2=0.25$
and the gravitational coupling $\alpha=0.5$.
The families of boson star solutions
are found in the $r_h \to 0$ limit.
They emerge from the vacuum at $\omega_{\rm max}=1$,
and form the fundamental branch of solutions.
Both the dependence of the mass and the
Noether charge of the regular configurations
on the frequency form an inspiraling pattern,
which is typical for spinning boson star solutions
and some other types of gravitating solitons
\cite{Kleihaus:2005me,Kleihaus:2007vk,Kleihaus:2015iea,Herdeiro:2018djx,Kunz:2013wka}.

As the fundamental branch of regular solutions
arises from the upper limiting frequency value $\omega_{\rm max}$,
the mass $M$ of the configurations gradually increases
with decreasing $\omega$
and,
for all values of the mass parameter $\mu$ not too close to zero,
the dependence $M(\omega)$ possesses a maximum at some
critical value of the frequency $\omega_{\rm M}$.
As the frequency decreases below that point,
the mass of the solutions decreases
until the minimum frequency $\omega_\mathrm{min}$ is reached.
Here the fundamental forward branch merges with a second (backward) branch,
leading to a counterclockwise inspiraling of the mass curve $M(\omega)$.

In general, for the same set of values of the parameters of the model,
the mass of the parity-odd configurations is
considerably higher than the mass of the parity-even boson stars,
as seen in Fig.~\ref{omega_rh_mu025}, left column.
The parity-odd $n=1$ solutions represent a new family of boson stars,
which in the flat space limit are linked to the corresponding Q-balls
\cite{Loiko:2018mhb}.

We have found that black holes with synchronized hair
exist in a small interval of the
event horizon radius parameter, typically $r_h < 0.07$,
see Fig.~\ref{omega_rh_mu025}.
For very small values of the horizon radius $r_h$, $r_h<0.02$, the inspiraling critical
behavior is changed to a multi-branch structure
with a few branches only, leading toward a second
upper critical value of the frequency $\omega_{cr}^{(1)}<1$,
see Fig.~\ref{omega_rh_mu025}, left plots.
In this limit, the real scalar component trivializes
and the lower branch ends on the vacuum Kerr black holes
with stationary scalar Klein-Gordon clouds,
see \cite{Hod:2012px,Herdeiro:2014goa,Herdeiro:2015gia,Herdeiro:2015waa,Herdeiro:2018djx,Herdeiro:2015tia}.

As the horizon radius $r_h$ increases, the multibranch structure is replaced by
a two-branch scenario with the second (lower, backward) branch
ending on the limiting Kerr solution as $\omega \to \omega_{cr}^{(1)}$,
see Fig.~\ref{omega_rh_mu025}, left column.
The maximum value of the frequency along the second branch
$\omega_{cr}^{(1)}$ slowly increases as the horizon radius $r_h$ grows,
and approaches $\omega_{\rm max}=1$ at some maximal value of the horizon radius $r_h$
as the loop shrinks to zero.
The branches exist up to the limit $\alpha=0$,
where the regular solutions approach the corresponding
flat space Q-ball configurations
while the solutions with non-zero horizon radius $r_h$ become linked
to the scalar clouds spinning around the Kerr black hole in the probe limit.

The Hawking temperature constantly decreases
as $\omega$ varies along the branches,
see Fig.~\ref{omega_rh_mu025}, middle column.
Note that the configurations with the smallest horizon radius $r_h$
have the smallest temperature,
both for the parity-even and parity-odd solutions.

Fig.~\ref{omega_rh_mu025} also presents the ADM mass
as a function of the angular momentum $J$, see the right column.
The mass of the regular spinning configurations with horizon radius $r_h=0$
exhibits a typical zig-zag behavior
known for boson stars
\cite{Friedberg:1986tp,Friedberg:1986tq,Kleihaus:2005me,Kleihaus:2007vk,Kleihaus:2015iea,Collodel:2017biu,Brihaye:2008cg}.
As the horizon radius increases,
it becomes replaced with the two-branch pattern,
which is well known for the Kerr scalar clouds and other similar solutions
\cite{Hod:2012px,Herdeiro:2014goa,Herdeiro:2015gia,Herdeiro:2015waa,Herdeiro:2018djx,Herdeiro:2015tia}.
Here the lower branch corresponds to the values of the frequency
smaller than $\omega_{cr}^{(1)}$,
while the upper branch represents the values $\omega > \omega_{cr}^{(1)}$.

In Fig.~\ref{fields_rh_omega_rh_mu025} we present the horizon
values of the massive scalar fields $X(r_h$ and $Y(r_h)$
as functions of the frequency $\omega$ for $\mu^2=0.25$.
Both the parity-even and the parity-odd solutions
start at $\omega_{\rm max}=1$ from the regular limits
$X(\omega_{\rm max})=1$, $Y(\omega_{\rm max})=0$.

To conclude the analysis of the massive ($\mu^2=0.25$)
spinning configurations,
we exhibit in Fig.~\ref{alpha} the dependence of the scalar mass of
the $n=1$ parity-even solutions on the gravitational coupling $\alpha$
for some set of the values of the horizon radius $r_h$.
One can see that the mass of the solutions on both branches
decreases monotonically as $\alpha$ increases.
In the limiting case $\alpha\to 0$ the solutions approach the Q-balls
spinning on the Minkowski or Schwarzschild background.

The situation changes drastically for the spinning solutions
when $\mu \to 0$.
First, we observe that the spinning component
of the coupled configuration falls off exponentially. It remains massive,
whereas the real scalar field decays as $\sim r^{-1}$.
Fig.~\ref{fields-profiles} exhibits the profile functions
of both parity-even and parity-odd scalar field functions
$X(r,\theta)$ and $Y(r,\theta)$ for the massive $\mu^2=0.25$
and the massless ($\mu=0$) case.
Thus, the massless limit provides a new type of hairy black hole with
hair of two different types, that are short- and long-ranged, respectively.

In Fig.~\ref{omega_rh_mu0} we display the ADM mass $M$ (left column)
and the Hawking temperature $T_H$ (middle column)
of the parity-even $n=1$ configurations versus the frequency $\omega$.
Similar to the solutions with non-zero values of the mass parameter $\mu$
presented in Fig.~\ref{omega_rh_mu025},
the fundamental branch of the regular spinning boson stars
arises in the limit $\omega_{\rm max} =1$,
and the mass of the configuration increases as the frequency decreases.
This branch extends up to some lower non-zero
critical value $\omega_\mathrm{min}>0$,
and the ADM mass of the boson star increases monotonically
along this branch.

The lower critical value $\omega_\mathrm{min}$
depends on the strength of the gravitational coupling,
and increases slowly as $\alpha$ grows.
In the flat space limit the axially-symmetric spinning Q-balls
with massless real component exist over the entire range
of values of the frequency $\omega\in [0,1]$
\cite{Levin:2010gp,Loiko:2018mhb}.
We constructed solutions with very large values of $\alpha$,
and they are likely to exist for arbitrary values
of the effective gravitational coupling.

\begin{figure}[hbt]
    \begin{center}
        \includegraphics[width=.3\textwidth, clip = true, trim = 150 150 150 150]{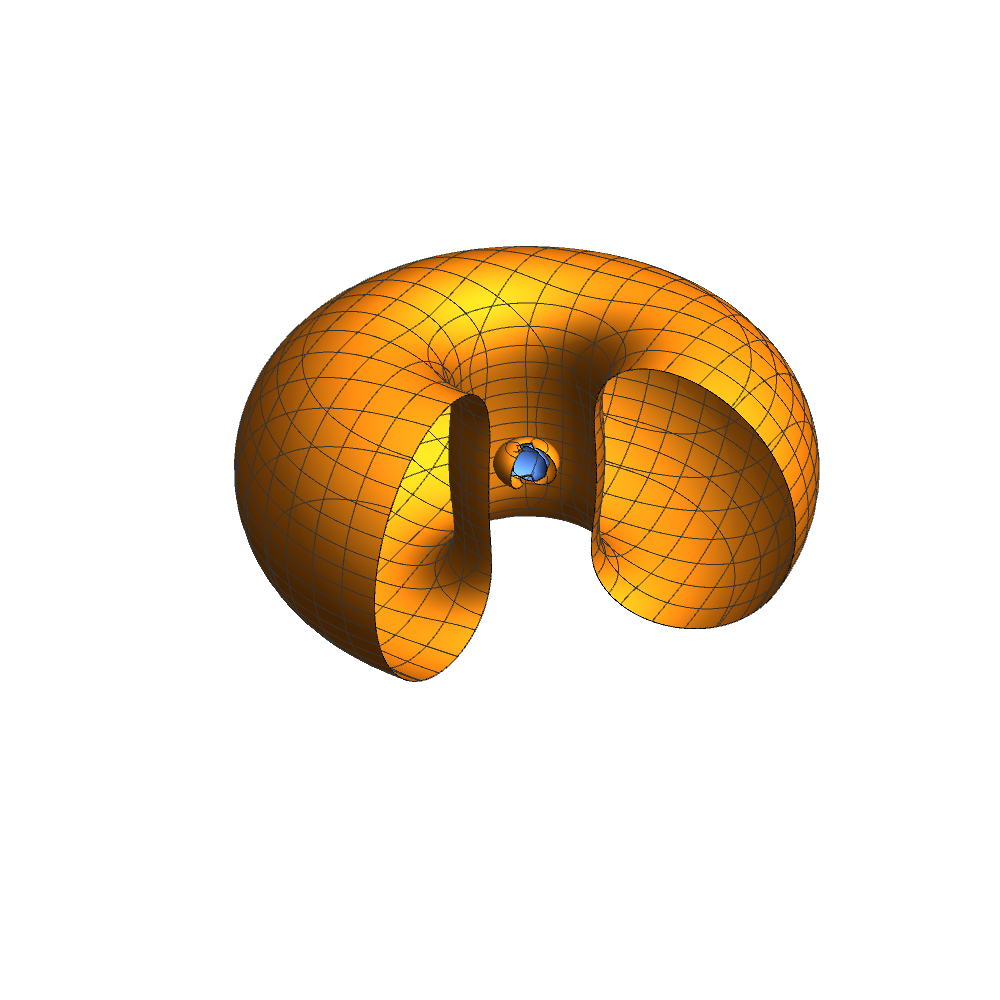}
        \includegraphics[width=.3\textwidth, clip = true, trim = 200 200 200 200]{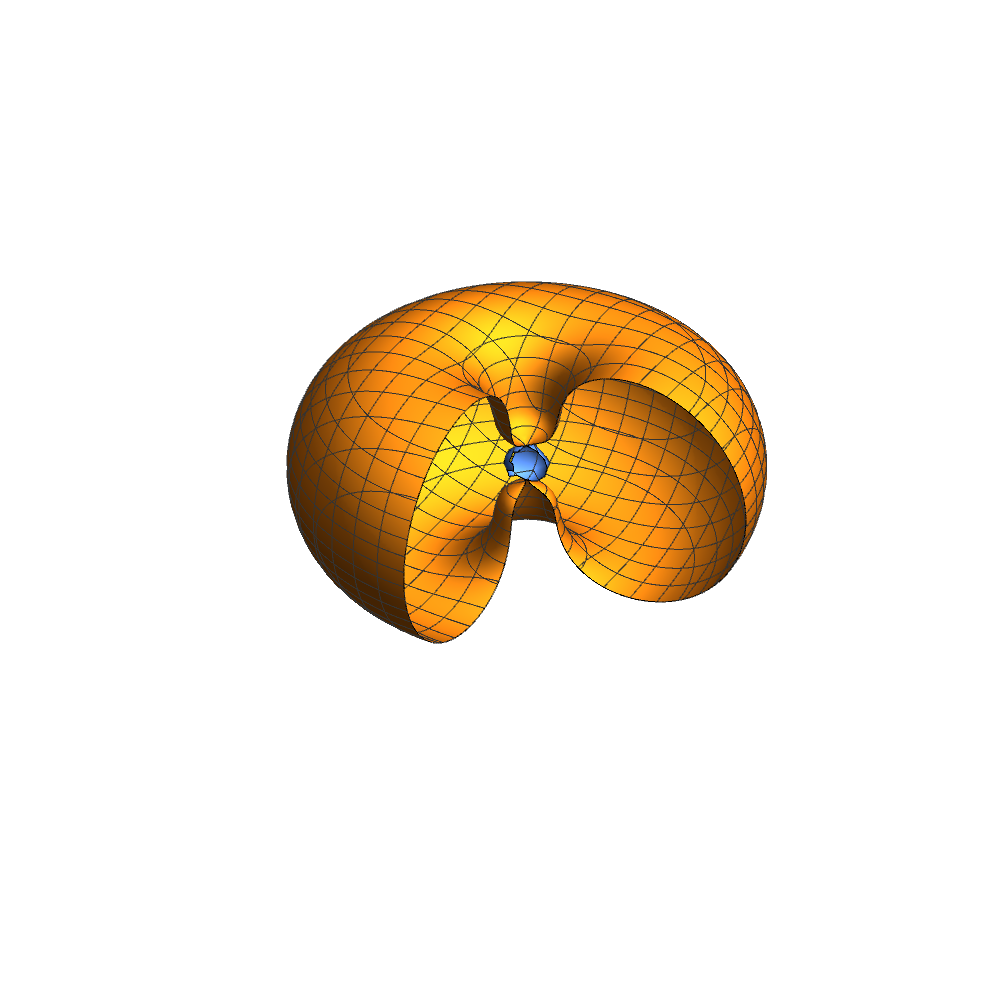}
        \includegraphics[width=.3\textwidth, clip = true, trim = 150 150 150 150]{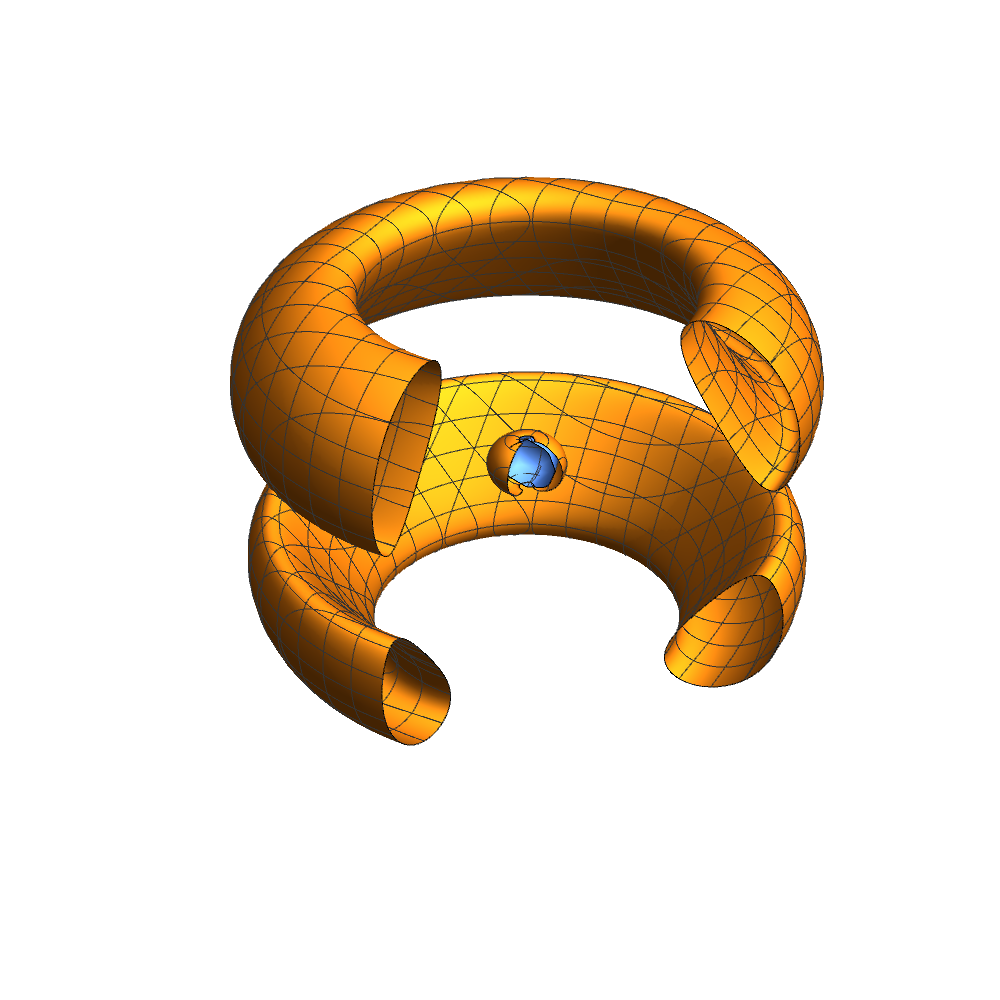}
    \end{center}
    \caption{\small
        Ergosurfaces of $n=1$ parity-even hairy BHs
        with horizon radius parameter $r_h=0.01$
        and frequencies $\omega/m=0.35$ (1), $\omega/m=0.5$ (2) and parity-odd BH with $\omega/m=0.5$ (3)
        on the backward branch. Blue surfaces represent the horizon.}
    \lbfig{ergo_bh}
\end{figure}

This branch bifurcates with the second upper branch at $\omega_\mathrm{min}$,
and we observe a small loop in the $M(\omega)$ dependence,
as illustrated in Fig.~\ref{omega_rh_mu0}, left plot,
see also the zoomed-in subplot in Fig.~\ref{parity}.
This loop, which is observed both for the parity-even
and parity-odd black holes with a synchronized massive component,
disappears when the mass parameter $\mu$ increases
sufficiently from zero.

The mass of the configurations rapidly increases
as the frequency $\omega$ approaches its minimal value $\omega_{min}$.
Further increase of the frequency $\omega$ along the second branch
is related with a decrease of the mass.
At some upper critical value of the frequency $\omega_{cr}^{(2)}$
the curve backbends and a third (backward) branch
of the regular spinning solitons is found.
Thus, an overall inspiral type pattern
is observed again for this sequence of solutions.

We have performed a similar study also for the parity-odd $\mu=0$ solutions
with a long-range real component, see Fig.~\ref{parity}.
The branch structure of the spinning parity-odd black holes with synchronized
hair is more explicit than the one of the corresponding parity-even solutions.
In particular, for small values of the horizon radius,
the third and the forth branch are clearly visible.

In addition, Fig.~\ref{omega_mu} exhibits the pattern
of the evolution of the branches of the parity-even (upper row)
and parity-odd (bottom row) solutions at the horizon radius $r_h=0.01$,
as the mass parameter $\mu$ varies.
We observe that the range of allowed values of the
frequency rapidly decreases as $\mu$ grows.
In the limiting case, $\mu\to \infty$,
the real scalar component becomes trivial everywhere in space, so
the model \re{action} becomes effectively truncated
to the massive complex-Klein-Gordon theory
minimally coupled to Einstein gravity \cite{Schunck:2003kk}.

It is known, that the latter model supports the existence of
black holes with synchronised hair
\cite{Hod:2012px,Herdeiro:2014goa,Herdeiro:2015gia}.
These solutions trivialize when the gravitational coupling is switched off.
Indeed, our simulations show
that the spinning component of the parity-even solutions
of the model \re{action} approaches the corresponding
solutions of the complex-Klein-Gordon theory presented in
\cite{Herdeiro:2014goa,Herdeiro:2015gia}.
On the other hand, in the limit $\mu\to \infty$
the $n=1$ parity-odd solutions of the model \re{lagFLS}
approach Kerr black holes with parity-odd synchronised scalar hair
\cite{Wang:2018xhw,Kunz:2019bhm}.

Finally, we address the geometry of the ergo-surfaces of the hairy black holes in the
Einstein-Friedberg-Lee-Sirlin model. The ergo-surfaces are
defined as the zero locus of the normalized time-like Killing vector $\xi \cdot \xi =0$, or
\be
g_{tt}= -F_0 + \sin^2 \theta  F_2 W^2 = 0 \, .
\label{ergo-sf}
\ee
We found that, in analogy to other known hairy black holes, there are both the
ordinary Kerr-like $S^2$ ergo-regions with topology $S^2$,
which appear on the fundamental branch, and ergo-Saturns  with
topology $S^2 \bigoplus (S^1\times S^1)$, when the black holes possess
parity-even scalar hair \cite{Herdeiro:2014jaa,Kleihaus:2015iea}, see Fig.~\ref{ergo_bh}.
Further, analogously to the corresponding parity-odd solutions in the Einstein-Klein-Gordon theory
\cite{Kunz:2019bhm}, there is a new type of ergo-surfaces, which
represent ergo-double-torus-Saturns with topology
$(S^1\times S^1)\bigoplus(S^1\times S^1)\bigoplus S^2$,
see Fig.~\ref{ergo_bh}, right plot.

\begin{figure}[hbt]
    \begin{center}
        \includegraphics[width=.48\textwidth, trim = 40 20 90 20, clip = true]{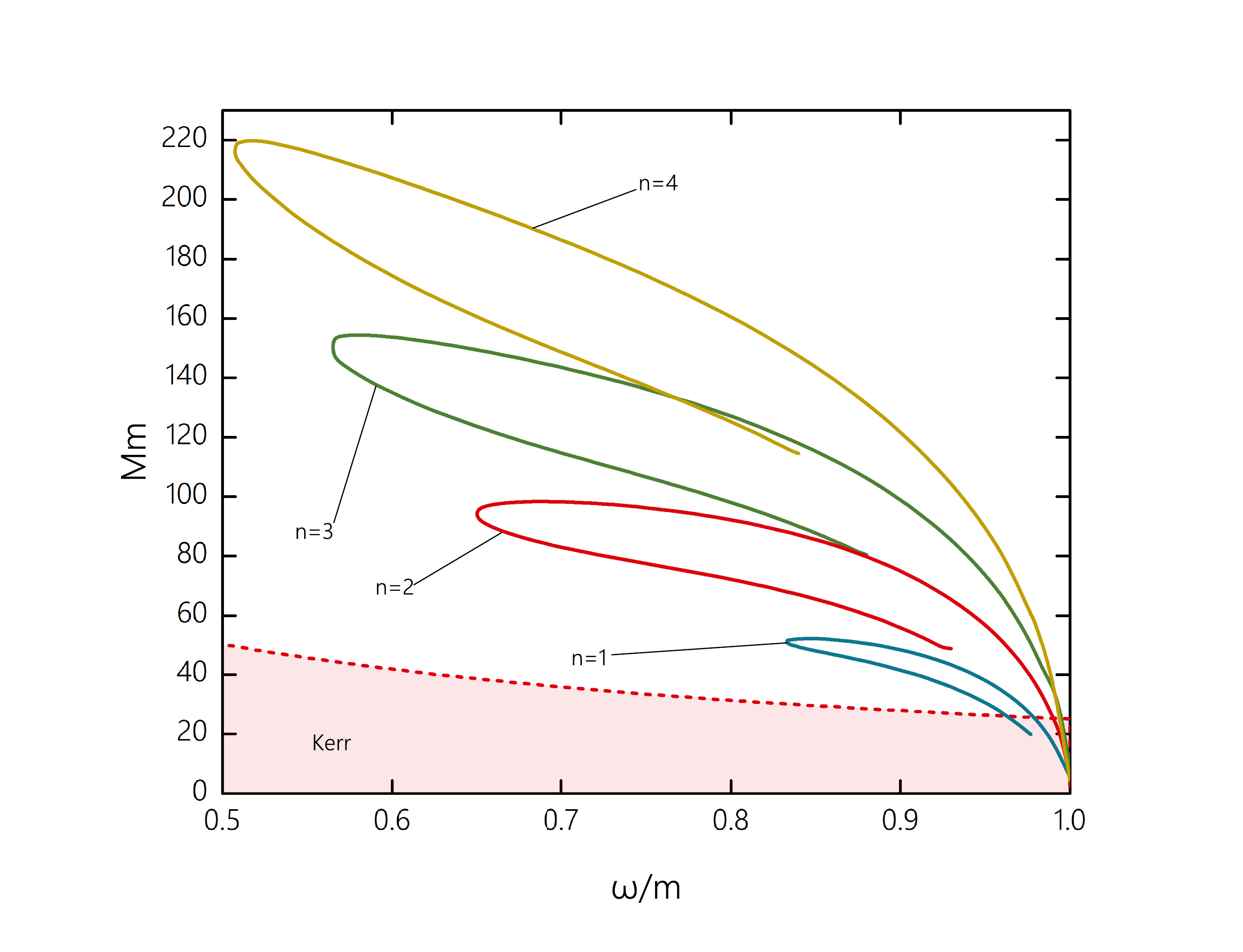}
        \includegraphics[width=.48\textwidth, trim = 40 20 90 20, clip = true]{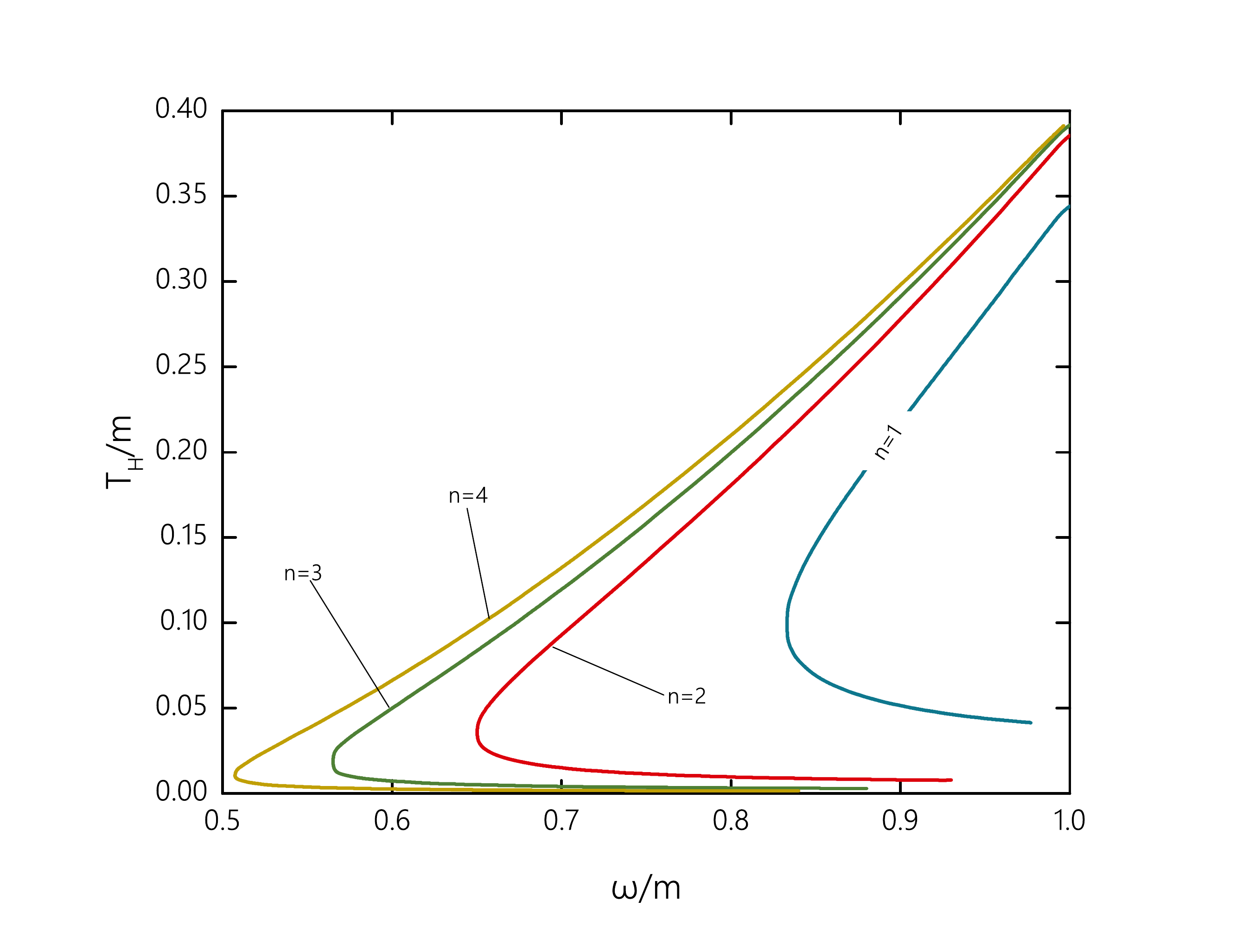}
    \end{center}
    \caption{\small
        The ADM mass $M$ (left plot)
and the Hawking temperature $T_h$ (right plot)
        vs the frequency $\omega/m$ for a set of values of
        the winding number $n$ for rotating parity-even
hairy black holes at the horizon radius $r_h=0.05$ and $\mu/m=0.25$.}
    \lbfig{omega_n}
\end{figure}

Also, we constructed spinning solutions for higher values
of the azimuthal winding number $n>1$.
Generally, these possess similar properties as the $n=1$ solutions,
however, their mass is higher, and they exist in a larger
interval of frequencies, as seen in Fig.~\ref{omega_n}.


\section{Conclusions}

In this work we have coupled the Friedberg-Lee-Sirlin model
\cite{Friedberg:1976me} to Einstein gravity
and investigated some of its solutions.
When the flat space non-topological solitons are coupled to
gravity, regular boson star solutions arise.
In the simplest case, these boson star solutions are
spherically symmetric. Here, however, we have concentrated
on their rotating generalizations, which may feature
a parity-even or parity-odd complex scalar field.

The properties of these boson stars are similar to those
found in models with a complex scalar field only,
like the Einstein-Klein-Gordon model.
In fact the solutions of this model are recovered,
when the coupling constant $\mu$ is taken to infinity.
On the other hand, also boson star solutions with a massless
real scalar field have been found.

When the presence of an event horizon
together with the synchronization condition between the
frequency and the horizon angular velocity is imposed,
new hairy black holes emerge in a certain region of
the parameter space.
According to the symmetries of the complex scalar field
they represent parity-even or parity-odd hairy black holes.

These hairy black holes possess two types of hair,
consisting of the usual complex scalar field hair
and the additional real scalar field hair.
In particular, when the mass of the real scalar field
vanishes, the real scalar field hair becomes
long-ranged, whereas the complex scalar field hair
remains massive. This constitutes an interesting
new quality of hairy black holes.

We note that similar hairy black hole solutions
as the ones studied here may also
exist in the Wick-Cutkosky model \cite{Wick-Cutkosky},
recently revisited in \cite{Nugaev:2016uqd,Panin:2018uoy},
if the corresponding Q-ball configurations
are coupled to Einstein gravity.

Various interesting features of the regular and hairy black hole solutions
of the Einstein-Friedberg-Lee-Sirlin model remain to be studied,
and, in particular, there should be numerous further
radially and angularly excited regular and hairy black hole solutions
in the model.

{\bf Acknowledgements}-- We are grateful to Burkhard Kleihaus and Eugen Radu for inspiring and valuable
discussions. This work was supported in part by the DFG Research Training Group 1620 {\sl Models of Gravity} as
well as by and the COST Action CA16104 {\sl GWverse}. Ya.S. gratefully acknowledges the support of the Alexander
von Humboldt Foundation and from the Ministry of Education and Science of Russian Federation, project No
3.1386.2017. I.P. would like to acknowledge support by the DAAD Ostpartnerschaft Programm.

\begin{small}

  \end{small}

\end{document}